\documentclass[aip,reprint]{revtex4-1}

\usepackage{amsmath,amsfonts,amssymb,mathrsfs,bm}
\numberwithin{equation}{section}
\usepackage{graphicx}

\usepackage{xcolor}
\usepackage[normalem]{ulem}
\newcommand{\f}{f}

\DeclareMathOperator{\tr}{\textrm{tr}}
\renewcommand{\vec}[1]{\boldsymbol{#1}}
\newcommand{\tens}[1]{\mathbf{#1}}
\newcommand{\p}{\partial}
\newcommand{\diff}{\textrm{d}}
\newcommand{\D}{\textrm{D}}
\newcommand{\rev}{\bigg|_{\textrm{rev.}} }
\newcommand{\Wi}{\textrm{Wi}}
\newcommand{\bnabla}{\vec{\nabla}}
\newcommand{\ens}[1]{\left\langle #1 \right\rangle}
\newcommand{\new}[1]{{\color{black}#1}}

\begin{document}
	
	
	\title{A thermodynamically consistent constitutive equation describing polymer disentanglement under flow} 
	
	
	
	\author{Benjamin E. Dolata}
	\affiliation{Georgetown University, Department of Physics and Institute for Soft Matter Synthesis and Metrology}
	
	\author{Peter D. Olmsted}
	
	
	\date{\today}
	
	\begin{abstract}
		We derive a thermodynamically consistent framework for incorporating entanglement dynamics into constitutive equations for flowing polymer melts. We use this to combine the convected constraint release (CCR) dynamics of Ianniruberto-Marriccui into a finitely-extensible version of the Rolie-Poly model, and also include  an anisotropic mobility as in the Giesekus model. The reversible dynamics are obtained from a free energy that describes both a finitely-extensible conformation tensor and an ideal gas of entanglements along the chain.  The dissipative dynamics give rise to coupled kinetic equations for the conformation tensor and entanglements, whose  coupling terms describe shear-induced disentanglement.  The relaxation dynamics of the conformation tensor follow the GLaMM and Rolie-Poly models, and account for reptation, retraction and CCR.  We propose that the relaxation time $\tau_\nu$ for entanglement recovery is proportional to the Rouse time $\tau_R$ which governs polymer stretch within the tube. This which matches recent molecular dynamics simulations, and corresponds to relaxing the entanglement number before the entire polymer anisotropy has relaxed on the longer reptation time  $\tau_d$. \new{Our model suggests that claimed signatures of slow re-entanglement on the reptation time in step-strain experiments may be  interpreted as arising from anisotropies in reptation dynamics.}
	\end{abstract}
	
	
	\maketitle 
	
	
	%
	%
	
	%
	
	
	
\section{Introduction}
Advances in polymer processing technologies require a microstructurally-aware viscoelastic constitutive equation that captures the stretch, orientation, and disentanglement of polymer melts under flow.  Polymer melts and solutions are often treated as generalized non-Newtonian fluids in traditional manufacturing processes such as single-screw extrusion\cite{cjse-96} because the relaxation time of the polymer is typically much smaller than \new{processing times}.  This separation of time scales no longer holds for additive-manufacturing techniques such as fused-filament fabrication, where both the residence time of the melt in the printer head and cooling time of the printed filaments are on the order of  the polymer's terminal relaxation time, causing the melt to undergo time-dependent stretch, orientation, and disentanglement.\cite{mo-17a,mo-17}  This behavior will influence the ultimate material properties of printed objects.  Molecular dynamics have shown that entangled melts stretch and orient with the flow, leading to disentanglement.\cite{maigm-03,fkmk-06,bmk-10,sek-15,sek-16,sek-19}   Indeed, at higher rates, some chains become completely disentangled, although a population of entangled chains remains.\cite{bmk-10}  Thus, accurate modeling of emerging manufacturing techniques requires constitutive equations that can capture the disentanglement of a melt under flow.

The  well-established tube model of entangled polymer melts is a mean-field model in which polymer molecules diffuse within an effective tube arising from confining constraints from the surrounding chains.\cite{DoiEdwards_Book}  This snake-like motion is termed reptation. The GLaMM model,\cite{glmm-03} a detailed refinement of the tube concept, accounts for convective constraint release (CCR),\cite{marrucci-96,im-96} contour-length fluctuations, and tube stretch.  However, tube models do not explicitly track the dynamics of entanglements.  A kink dynamics algorithm develop by \citet{msl-19} explicitly models entanglements.  However, this model predicts complete disentanglement of the melt under flow, in contradiction with simulations.\cite{maigm-03,fkmk-06,bmk-10,sek-15,sek-16,sek-19a,sek-19}   The slip-link model overcomes these limitations by explicitly modeling the creation, destruction, and motion of entanglements.\cite{schieber-03,ks-09,jks-12,sis-13}  In this model, entanglements on a test chain arise from equilibration with a chemical potential bath, and may \new{be} lost via CCR.  The discrete slip-link model is equivalent to a continuous tube model in highly-entangled melts.\cite{ss-12}

\new{The GLaMM and slip-link models can accurately predict the rheology response of polymer melts over a wide range of timescales because they capture melt dynamics on the entanglement time $\tau_e$, which is the Rouse time of a tube segments between entanglements.  However, such detailed modeling inherently requires many degrees of freedom to describe the multiple tube segments in each chain, and the computational requirements associated with these degrees of freedom render them unsuitable for simulating the complex flows of manufacturing processes.  It would therefore be desirable to obtain simple closed-form constitutive equations with fewer degrees of freedom that are still capable of describing the stretch, orientation, and disentanglement of a flowing polymer melt.  Ideally, such a  constitutive equation should describe all polymer dynamics relevant on processing time scales.  Typical processing flows have shear rates much smaller than $1/\tau_e$, and residences times much larger than $\tau_e$.\cite{mo-17}  Thus, the individual tube segments will fluctuate rapidly and sample their equilibrium distribution on timescales comparable to the flow timescales. 

This implies that these rapid fluctuations can be coarse-grained out, and that only slower dynamics on the Rouse time $\tau_R$ and reptation time $\tau_d$ are relevant.}
The contour length of a polymer fluctuates within the tube on the Rouse time $\tau_R$ of the entire chain.\cite{DoiEdwards_Book} Such fluctuations lead to retraction along the tube on the Rouse time.\cite{mld-98,glmm-03,lg-03} The chain escapes the original confining tube on the longer reptation time $\tau_d$, forming a new set of entanglements.  In the Doi-Edwards model the reptation time is the time it takes a fixed-length chain to escape the tube by curvilinear diffusion.\cite{DoiEdwards_Book} In practice the reptation time is reduced from the Doi-Edwards prediction due to contour-length fluctuations\cite{DoiEdwards_Book,lm-02} and CCR events\cite{marrucci-96,mld-98,glmm-03,lg-03} on the Rouse time.

\citeauthor{mld-98} (MLD) developed a simple constitutive equation that accounts for reptation, retraction, and CCR, which  predicts the stretch and orientation of polymers in  flow\cite{mld-98}.  While the MLD model successfully predicts many experimental observations in steady and transient shear flow, the proposed decoupling of stretch and orientation is less numerically stable than a single conformation tensor describing both stretch and orientation\cite{wk-04}.  The Rolie-Poly model,\cite{lg-03} obtained via simplification of the GLaMM model,  overcomes this limitation by describing stretch and orientation with a single conformation tensor variable.  While the model describes chain reptation, retraction, and CCR, it does not explicitly track the evolution of the number of entanglement.

There have been a few attempts to develop constitutive equations that explicitly compute the dynamics of entanglements. The earliest were based on rubber elasticity.\cite{ygm-97}  Subsequent models assume that the melt disentangles via CCR at a rate proportional to the non-affine stretch rate of the molecules.  The first of such models, due to \citeauthor{im-14}, \new{assumes} that entanglements are lost via CCR and regained by reptation.\cite{im-14,im-14b,ianniruberto-15}  A nearly identical model of disentanglement was obtained independently by \citet{hhhr-15} for branched polymers.  Finite extensibility was incorporated for monodisperse \cite{mbp-15} and  polydisperse melts.\cite{mmp-18}  While the  Ianniruberto-Marrucci (I-M) model\cite{im-14,im-14b,ianniruberto-15}  produces good agreement with molecular dynamics simulations of flow-induced disentanglement,\cite{im-14,bmk-10} it possesses three drawbacks.  First, the stress in the model is expressed as a history integral over all past times, making it unsuitable for use in large-scale simulations.  Second, the relaxation mechanics assumes re-entanglement on the reptation time, which
contradicts observations from recent molecular dynamics simulations of \new{Kramer-Grest\cite{ohr-19,Marco_Thesis} and united-atom polyethelyne\cite{bsek-22} melts.} Finally, we show below that the model is not consistent with the Onsager-Casimir reciprocal relations, required within the generalized transport matrix governing coupled relaxations.\cite{onsager-31a,onsager-31b,casimir-45,BerisEdwards,Ottinger_Book}

The primary goal of this work is to derive a thermodynamically-consistent constitutive equation predicting flow-induced disentanglement of a polymer melt.  We express the convective constraint release mechanism employed in prior works in a thermodynamically consistent form and derive a re-entanglement mechanism consistent with the unexpectedly fast re-entanglement observed in recent molecular dynamics simulations.\cite{ohr-19,Marco_Thesis,bsek-22}  The remainder of the paper is organized as follows.  In Sec.~\ref{sec:theoretical-background}, the kinematics of an entangled melt are interrogated, and these results are used to inform the theoretical background of the Ianniruberto-Marrucci model in Sec.~\ref{sec:im-model}.  We then use the results of the prior two sections to derive a thermodynamically consistent framework in Sec.~\ref{sec:thermodynamic}.  We \new{obtain} a physically motivated expression for the free energy of entanglements, which we use to derive thermodynamically consistent evolution equations for the melt conformation and entanglements using the single-generator formulation\cite{BerisEdwards} of non-equilibrium thermodynamics.  We apply this to derive a specific constitutive model that blends the physics inherent in the I-M disentanglement mechanism,\cite{im-14,im-14b,ianniruberto-15} the Rolie-Poly model, \cite{lg-03} and the Giesekus model.\cite{giesekus-82} Some analytical results for  the model are presented in Sec.~\ref{sec:analytical}. \new{We obtain numerical solutions of our model for disentanglement under flow in Sec~\ref{sec:results}.  We show that the disentanglement dynamics are consistent with molecular dynamics simulations, and that the rheological predictions of our model are consistent with experiments.}

\section{Kinematics of an entangled melt}\label{sec:theoretical-background}
In this section, we define the structural variables we employ to parameterize the melt and derive their kinematics.  We begin in Sec.~\ref{subsec:conformation} by defining a conformation tensor that parameterizes the stress, as well as its kinematics under reversible flows which will later be used to derive the stress tensor.  We then derive a relationship between stretch, orientation, and conformation in Sec.~\ref{subsec:stretch-orientation}, which will ultimately allow us to determine the disentanglement mechanism.

\subsection{Melt conformation and entanglements}\label{subsec:conformation}

\begin{figure}
	\raggedleft
	\begin{minipage}[b]{.115\textwidth}
		(a) 
		\vspace{60pt}
	\end{minipage}
	\includegraphics[height=90pt]{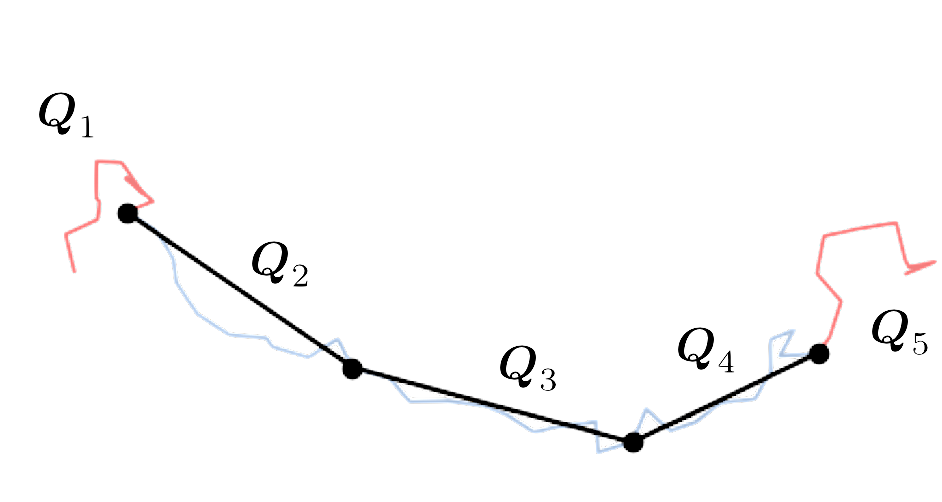}\\
	\begin{minipage}[b]{.02\textwidth}
		(b) 
		\vspace{60pt}
	\end{minipage}
	\includegraphics[height=90pt]{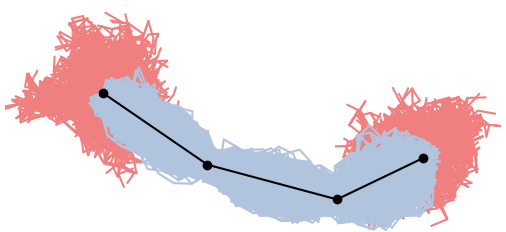}
	\caption{(a: Representation of a chain in an entangled melt, black lines represent the primitive path while blue and red lines are the polymer chain; red corresponds to the dangling ends.  The black circles represent the $Z_e=4$ entanglements.  (b) The constraining tube becomes evident after the chain explores many configurations. }\label{fig:tube}
\end{figure}
We parameterize the melt with a tube conformation tensor encoding the stretch and orientation of the polymer tube.  The tube comprises $Z_e+1$ segments connected by $Z_e$ entanglements.  Entanglements are topological constraints on the motion of the polymer chain.  The chain only feels  entanglements on timescales comparable to or larger than the Rouse time $\tau_e$ of an entanglement strand, during which period the chain samples many configurations for a `tube' to appear (\textit{c.f.} figure~\ref{fig:tube}).  \new{We assume that the topological entanglements $Z_e$ are equivalent to the kinks measured in chain-shrinking algorithms such as the Z1 code.\cite{kroger-05}.  Such coarse-graining algorithms cannot incorporate all aspects of polymer interactions.  However, the kink measurements in the Z1 code are predictive of material strength,\cite{gppgr-13,cr-20} suggesting that the Z1 code captures the interactions that are most relevant to macrcosopic material properties of polymer melts.}

\new{In simulations, the number of kinks, which we identify with the number of topological entanglements $Z_e$, is distinct from the chain-shrinking algorithms can be used to measure the number of rheological tube segments $Z_\textrm{rheol}$ computed from the tube diameter. Generally, $Z_\textrm{rheol}$ about half the value of $Z_e$ at equilibrium.\cite{maigm-03,tt-06,fkmk-06,bmk-10}  This difference is indicative of correlations among the alignment of neighboring  tube segments.\cite{tt-06}  In slip-spring models, such correlations lead to non-affine displacements in the melt in response to an applied shear, and hence a plateau modulus smaller than the value of $(Z_{e,eq}+1)nk_\textrm{B}T$ expected from the equipartition theorem.\cite{sis-13}}

The tube conformation tensor has the form (Appendix~\ref{app:conformation-tensor})
\begin{equation}\label{eq:conformation-tensor-definition}
\tens{A} = \frac{3}{b_K^2(Z_{e,eq} + 1)}\ens{\int_0^{Z_e+1}\frac{\vec{Q}(s)\vec{Q}(s)}{N_e(s)} \diff s},
\end{equation}
where we use a continuous representation of the tube and $Z_{e,eq}$ is the average number of entanglements at equilibrium.  The prefactors ensure that $\tens{A}=\tens{I}$ at thermodynamic equilibrium.\cite{schieber-03}  Here, $s$ is a dimensionless tube coordinate, and $\vec{Q}(s)$ and $N_e(s)$ are the tube segment end-to-end vectors and number of Kuhn steps in that tube segment.  The tube-segment vectors can alternatively be represented as $\diff\vec{R}(s)/\diff s$, where $\vec{R}(s)$ is the position of the center of tube segment $i$.  This conformation tensor defined in \eqref{eq:conformation-tensor-definition} is proportional to the stress, and is consistent with the stress optical rule.\cite{ss-12,sis-13} 

\new{The conformation tensor encodes the stretch and orientation of the melt.  \new{The} normalized entanglement number 
\begin{equation}
	\nu = \frac{\ens{Z_e}}{Z_{e,eq}}
\end{equation}
\new{parametrizes} the number of entanglements.}

\subsection{Stretch and orientation}\label{subsec:stretch-orientation}
The Ianniruberto-Marrucci model\cite{im-14,im-14b,ianniruberto-15} assumes that the melt disentangles due to non-affine stretch.  The physical origin of this non-affine deformation is chain relaxation; entanglements are lost at the chain ends as the chain retracts.  Although the definition (Eq.~\ref{eq:conformation-tensor-definition}) of the conformation tensor is generally agreed upon, there are many different definitions of stretch and orientation in the literature. Here, we give a general relationship between stretch, orientation, and the conformation tensor, \new{which unifies} these definitions into a general framework. These results will be employed in Sec.~\ref{sec:thermodynamic} to construct a thermodynamically consistent mechanism of stretch-induced disentanglement.

In its simplest definition, the stretch is proportional to the length of the tube segments i.e. $\lambda_T\sim \ens{|\vec{Q}|}$.  In this case the stretch is the normalized arc-length of the tube:
\begin{equation}\label{eq:stretch-tube}
	\lambda_T = \frac{1}{L_{eq}}\ens{\int_0^{Z_e+1}|\vec{Q}(s)|\diff s},
\end{equation}
where the normalization by the equilibrium tube length $L_{eq}\sim\sqrt{(Z_{e,eq}+1)}N_Kb_K$\cite{schieber-03,ks-08} ensures  that $\lambda_T=1$ at equilibrium.  While the tube stretch \eqref{eq:stretch-tube} is well defined, it does not have a clear relation to the conformation tensor and hence the stress.  An alternative definition can be found by recognizing that $\tr\tens{A}\sim\ens{|\vec{Q}|^2}$, and hence $\sqrt{\tr\tens{A}/3}$ is the mean-squared stretch of the melt, allowing us to define the stretch and orientation tensor as in, \textit{e.g.}, the Rolie-Poly equation:\cite{lg-03}
\begin{subequations}\label{eq:stretch-orientation}
	\begin{align}
		\tens{A} & = 3\lambda_{MS}^2\tens{S}_{MS}, \label{eq:stretch-orientation-decoubling} \\
		\lambda_{MS} & = \sqrt{\frac{\tr\tens{A}}{3}}, \label{eq:stretch}\\
		\tens{S}_{MS} & = \frac{\tens{A}}{\tr\tens{A}}, \label{eq:orientation}
	\end{align}
\end{subequations}
where the subscript $MS$ stands for mean-squared.  We show in Appendix~\ref{app:tube-stretch} that
\begin{equation}
0.86\lambda_{MS}\leq\lambda_T\leq\lambda_{MS}.
\end{equation}
The lower bound applies for fully aligned systems, while an isotropic conformation tensor corresponds to $\lambda_T=\lambda_{MS}$. We will use $\lambda_{MS}$ to model the retraction and CCR dynamics in our constitutive model (\textit{c.f.} Sec.~\ref{sec:thermodynamic}).

 We show in Appendix~\ref{app:conformation-entanglements-kinematics} that these two definitions ($\lambda_{MS}$ and $\lambda_T$) can be unified through a general relationship among the stretch, the orientation tensor $\tens{S}$, and the conformation tensor:
\begin{equation}\label{eq:stretch-orientation-equation}
	\tens{S} = 2\tens{A}\cdot\frac{\p \ln\lambda}{\p \tens{A}}.
\end{equation}
This expression automatically satisfies the constraint $\tr\tens{S}=1$.  Here, $\lambda$ and $\tens{S}$ are generic and do not represent  particular definitions of stretch and orientation [for example, $\tens{S}=\tens{A}^q/\tr\tens{A}^q$ is a general orientation tensor, with an associated stretch $\lambda_q=(\tr\tens{A}^q/3)^{1/q}$].\cite{ianniruberto-15}  This relationship between the stretch and orientation tensor is not completely general; one might define a stretch relative to a non-equilibrium tube as was done by \citeauthor{mbp-15}.\cite{mbp-15} However, \eqref{eq:stretch-orientation-equation} is consistent with most definitions of stretch and orientation in the literature, including the rigorous Doi-Edwards stretch and orientation\cite{DoiEdwards_Book}, the logarithmic stretch and the independent alignment approximation for the orientation tensor,\cite{wehm-98} and the deformation measures employed in the I-M model.\cite{ianniruberto-15}.

\section{Flow induced disentanglement: The Ianniruberto-Marrucci model}\label{sec:im-model}
\citet{im-14,im-14b} \cite{ianniruberto-15} obtained a tube-model informed kinetic equation for disentanglement
\begin{equation}\label{eq:im-model}
	\frac{\D \nu}{\D t} = -\frac{\beta\nu}{\lambda}\left(\lambda\tens{S}:\bnabla\vec{v} - \frac{\D \lambda}{\D t} \right) - \frac{\nu - 1}{\tau_{d,eq}}.
\end{equation}
Here, $\beta$ controls the rate of convective constraint release, and is inversely proportional to the number of retraction events necessary to release one entanglement.  The equilibrium reptation time $\tau_{d,eq}$, \cite{lm-02}
\begin{equation}\label{eq:tau-d-eq}
	\frac{\tau_{d,eq}}{3Z_\textrm{rheol}} = \left(1 - \frac{3.38}{Z_\textrm{rheol}^{1/2}} + \frac{4.17}{Z_\textrm{rheol}} - \frac{1.55}{Z_\textrm{rheol}^{3/2}} \right)\tau_R.
\end{equation}
is the time for a polymer to escape its tube \new{and} assume an isotropic configuration\new{,} thus relaxing its stress. The Rouse time
\begin{equation}
	\tau_R=Z_\textrm{rheol}^2\tau_e
\end{equation}
governs the timescale of retraction and contour-length fluctuations within the tube, and $\tau_e$ is the \new{entanglement time, \textit{i.e.} the} (Rouse) relaxation time of a single tube segment.  The number of steps in the rheological primitive path is $Z_\textrm{rheol}$, and is determined from the plateau modulus as discussed below.  The expression for the reptation time includes the influence of contour-length fluctuations, and approaches the Doi-Edwards value $\tau^{\textrm{DE}}_{d,eq}=3Z_\textrm{rheol}\tau_R$ in the limit of large $Z_\textrm{rheol}$.

The first term  in \eqref{eq:im-model} arises from the assumption that the rate of disentanglement is proportional to the non-affine stretch rate of the melt.  Re-entanglement towards equilibrium ($\nu=1$) is assumed to occur as a linear relaxation process governed by the reptation time, in common with other models of flow-induced disentanglement.\cite{hhhr-15,mbp-15}  This assumption is contradicted by molecular dynamics simulations,\cite{ohr-19,Marco_Thesis,bsek-22} which find that polymer chains relax back to the equilibrium tube on some re-entanglement time $\tau_\nu\ll\tau_{d,eq}$ rather than the reptation time $\tau_{d,eq}$.  This contradiction will be revisited in Sec.~\ref{subsec:tau-nu}.

The stress tensor in the I-M model is given by 
\begin{equation}\label{eq:stress-im}
\vec{\sigma} = 9G_0\lambda^2\tens{S},
\end{equation}
with  plateau modulus
\begin{equation}
G_0 = \rho RT/M_{e,eq},
\end{equation}
where $\rho$ is the melt density, $R$ is the gas constant, $T$ is the temperature, and $M_{e,eq}$ is the equilibrium entanglement molecular weight.  The factor of nine arises due to the specific form of $\tens{S}$ used by I-M, and ensures that the viscosity is equal to $G_0\tau_{d,eq}$ in the linear response regime.  Detailed expressions for stretch and orientation in the I-M model are given in Appendix~\ref{app:im}.

An equivalent expression for the plateau modulus is
\begin{equation}
G_0 = Z_\textrm{rheol}nk_\textrm{B}T,
\end{equation}
where $Z_\textrm{rheol}$ is the number of rheological tube segments, $n$ is the number density of polymer chains, and $k_\textrm{B}$  is Boltzmann's constant.  The number of segments in the rheological tube is defined by the entanglement molecular weight
\begin{equation}
Z_\textrm{rheol} = \frac{M}{M_{e,eq}},
\end{equation}
where $M$ is the polymer's molecular weight. \new{As discussed in Sec. \ref{subsec:conformation}, $Z_\textrm{rheol.}$ is generally not equal to $Z_e$ due to correlations between tube segments.} 


The modulus in \eqref{eq:stress-im} does not explicitly depend on the number of entanglements, in apparent contradiction with the assumption of an entanglement-dependent modulus 
\begin{equation}
	G(\nu) = G_0\frac{Z_{e,eq}\nu + 1}{Z_{e,eq} + 1} 
\end{equation}
made in some prior studies.\cite{ygm-97,hhhr-15,mbp-15,sek-19}  This apparent contradiction can be resolved from careful consideration of definitions of the stretch.  The conformation \new{tensor} in Sec.~\ref{subsec:conformation} is defined relative to its equilibrium value.  As shown in \eqref{eq:conformation-tensor-transformation}, this conformation tensor includes the influence of disentanglement in its definition.  This same approach is followed by \citeauthor{im-14} in their definition of the stretch, \new{which implies} a constant plateau modulus.\cite{im-14b}  On the other hand, \citet{mbp-15} defined a stretch $\Lambda$ relative to a non-equilibrium tube with $Z_{e,eq}\nu$ entanglements, which in our notation reads
\begin{equation}
	\Lambda = \sqrt{\frac{Z_{e,eq}\nu + 1}{Z_{e,eq} + 1}}\lambda.
\end{equation}
Hence, $G(\nu)\Lambda^2=G_0\lambda^2$.  Thus, both definitions of the modulus are equivalent, and simply refer to different definitions of the stretch.

The I-M disentanglement mechanism \eqref{eq:im-model} may now be expressed in terms of the structural variables described in \eqref{eq:stretch-orientation-decoubling}, yielding
\begin{equation}
\frac{\D \nu}{\D t} = -\frac{\beta\nu}{\tr\tens{A}}\left(\tens{A}:\bnabla\vec{v} - \frac{1}{2}\frac{D \tr\tens{A}}{D t}\right)	-
\frac{\nu - 1}{\tau_{d,eq}}.
\end{equation}
A different expression was employed by \citet{mo-17a,mo-17}:
\begin{equation}\label{eq:im-mo}
\frac{\D \nu}{\D t} = -\beta\nu\left(\tens{A}:\bnabla\vec{v} - \frac{1}{\tr \tens{A}}\frac{\D \tr\tens{A}}{\D t}\right)	-
\frac{\nu - 1}{\tau_{d,eq}}.
\end{equation}
This expression assumes that the rate of disentanglement is proportional to the conformation tensor $\tens{A}$, rather than the orientation tensor $\tens{S}=\tens{A}/\tr\tens{A}$.  At high shear rates, the eigenvalues of $\tens{A}$ approach the maximum polymer stretch, while the eigenvalues of $\tens{S}$, by definition, cannot exceed $1$.  As a consequence, \eqref{eq:im-mo}  overpredicts disentanglement at high shear-rates.  

\new{Physical insight can be obtained by using the chain rule to express the orientation tensor and the material derivative of the stretch in the \citeauthor{im-14} model \eqref{eq:im-model} in terms of $\tens{A}$:}
\begin{equation}
    \begin{split}
	\frac{\D \nu}{\D t} & = -\frac{\beta\nu}{\lambda}\left[2\frac{\p\lambda}{\p \tens{A}}:\left(\tens{A}\cdot\bnabla\vec{v}\right) - \frac{\D \tens{A}}{\D t}:\frac{\p \lambda}{\p \tens{A} }\right] \\
	&\quad  - \frac{\nu - 1}{\tau_{d,eq}}.
    \end{split}
\end{equation}
We recognize the term in square brackets as the upper-convected derivative, yielding
\begin{equation}\label{eq:im-upper-convected}
	\frac{\D \nu}{\D t} = \frac{\beta\nu}{\lambda} \frac{\p \lambda}{\p \tens{A}}:\stackrel{\triangledown}{\tens{A}}  - \frac{\nu - 1}{\tau_{d,eq}}.
\end{equation}
The first term projects the non-affine deformation $\stackrel{\triangledown}{\tens{A}}$ along the stretching direction specified by  $\p\lambda/\p\tens{A}$.  Physically, this captures the relaxation of the melt relative to the material deformation in the direction the tube is stretching.

\section{Thermodynamically consistent disentanglement model}\label{sec:thermodynamic}
The kinetic equation for $\nu(t)$ in the Ianniruberto-Marrucci model (Eq.~\ref{eq:im-model}) can be used to compute disentanglement in steady-state shear flow, producing good agreement with the simulations of disentanglement in steady-state shear flows.\cite{bmk-10}  However, there are at least two drawbacks with this model. 
First, we will show below that such a model requires a companion term in the equation of motion for the conformation tensor in order to satisfy the Onsager-Casimir reciprocal relations that guarantee positive entropy production.\cite{onsager-31a,onsager-31b,casimir-45}  Second, the assumption of re-entanglement on the reptation time contradicts the faster re-entanglement observed in simulations.\cite{ohr-19,Marco_Thesis,bsek-22}  We thus seek a thermodynamically consistent model that incorporates the same physics as  the I-M model, together with a re-entanglement time $\tau_\nu$ that is consistent with simulations.  We begin by deriving an expression for the free energy of an entangled melt in Sec.~\ref{subsec:free-energy}, followed by the construction of a framework for the dissipative dynamics of the melt in Sec.~\ref{subsec:evolution-equations}. The consequence of Onsager-Casimir reciprocity on the coupling between stretch and disentanglement is discussed in Sec.~\ref{subsec:onsager-casimir}. \new{The explicit forms of the dissipative dynamics of the conformation tensor and normalized entanglement number are determined in Sec.~\ref{subsec:mobility}, the full governing equations are presented in Sec.~\ref{subsec:governing-equation}, and the timescale of re-entanglement is determined in Sec.~\ref{subsec:tau-nu}.}

\subsection{Free energy of an entangled melt}\label{subsec:free-energy}
The stress follows from the Helmholtz free energy $F=F_A + F_\nu$, which we assume has separate contributions $F_A$ and $F_\nu$, respectively arising from the conformation tensor and entanglements.  We employ a finitely extensible non-linear elastic (FENE) form of the free energy for the conformation tensor
\begin{equation}\label{eq:free-energy-dimensional-A}
    F_A = \int \frac{ G_0}{2}\left(\phi\left(\lambda\right)
        -  \ln \det\tens{A}\right) \diff^3 r,
\end{equation}
where $\phi(\lambda)$ is the FENE-potential of the melt.  Henceforth, we define the stretch as the mean-squared stretch $\lambda\equiv\lambda_{MS}=\sqrt{\tr\tens{A}/3}$, as in the Rolie-Poly equation.\cite{lg-03}   The contribution to the free energy from the conformation tensor is independent of entanglements, which means that the stretched melt will relax back towards the equilibrium tube rather than to a non-equilibrium tube determined by the current number of entanglements. This is incompatible with the original slip-link model,\cite{schieber-03} which predicts that the stretch will relax to a non-equilibrium tube set by the current number of entanglements; but it is consistent with molecular dynamics simulations \new{of re-entanglement following cessation of an extensional flow} that \new{observed} fast equilibration of the tube diameter on the Rouse time\cite{ohr-19},  \new{and exponential recovery of entanglements on the order of the Rouse time.\cite{Marco_Thesis,bsek-22}} We employ the Cohen approximation\cite{cohen-91,sbm-09}
\begin{equation}\label{eq:phi}
	\begin{aligned}
		\phi(\lambda(\tens{A})) = \lambda^2 - 1 - 2(\lambda_\textrm{max}^2 - 1)\ln\frac{\lambda_\textrm{max}^2 - \lambda^2}{\lambda_\textrm{max}^2-1} , \\
			= \frac{\tr\tens{A}-3}{3} - 2(\lambda_\textrm{max}^2 - 1)\ln\frac{3\lambda_\textrm{max}^2 -\tr\tens{A}}{3(\lambda_\textrm{max}^2 - 1)}
	\end{aligned}
\end{equation}
for the FENE potential, where $\lambda_\textrm{max}=L_\textrm{max}/L_{eq}$ is the maximum stretch.  The maximum stretch can be expressed as
\begin{equation}\label{eq:maximum-stretch}
	\lambda_\textrm{max}^2 = \frac{1}{b_K^2(Z_{e,eq} + 1)}\ens{\int_0^{1+Z_e}\frac{\vec{Q}_\textrm{max}\cdot \vec{Q}_\textrm{max}}{N_e(s)}\diff s}.
\end{equation}
using \eqref{eq:conformation-tensor-definition}, where maximum extension of a tube segment is $|\vec{Q}_\textrm{max}| \sim b_KN_e(s)$.  The integrand becomes $b_K^2N_e(s)$ after substitution of the maximum stretch.  The integral of $N_e(s)$ (the number of Kuhn steps per segment) over the entire tube is the total number of Kuhn steps $N_K$, so that
\begin{equation}\label{eq:maximum-stretch-derived}
	\begin{split}
		\lambda_\textrm{max} & = \sqrt{\frac{N_K}{(Z_{e,eq} + 1)}}, \\
			& = \sqrt{N_{e,eq}},
	\end{split}
\end{equation}
which is independent of the number of entanglements.

To estimate the  entanglement contribution to the free energy we first  observe that molecular dynamics simulations show a Poisson distribution of entanglements, even when very far from equilibrium.\cite{fkmk-06,bmk-10}  We therefore approximate the entanglement contribution to the free energy as the Gibbs entropy $S=-k_\textrm{B}\sum_ip_i\ln p_i/p_{i,eq}$ for a Poisson distribution $p_i$, leading to the Helmholtz free energy
\begin{equation}
F_\nu =	Z_{e,eq}nk_BT\int  \nu\left(\ln\nu  - 1\right)\diff^3 r.
\end{equation}
This can also be characterized as the entropy of an ideal gas of entanglements distributed along the tube.  The prefactor is related to the plateau modulus through the expression
\begin{equation}
Z_{e,eq}nk_BT = \zeta_ZG_0,
\end{equation}
where
\begin{equation}
\zeta_Z = \frac{Z_{e,eq}}{Z_\textrm{rheol}}.
\end{equation}
Throughout this work, we employ
\begin{equation}
	\zeta_Z = \frac{2}{1 + Z_{e,eq}^{-1}},
\end{equation}
which follows from molecular dynamics simulations\cite{maigm-03,tt-06,fkmk-06,bmk-10} as discussed in \new{Sec.~\ref{subsec:conformation}}.  Near equilibrium $(\nu=1)$ the ideal gas free energy is approximated as
\begin{equation}
\zeta_Z G_0 \nu\left(\ln\nu  - 1\right) \simeq \frac{\zeta_Z G_0 }{2}\left((\nu - 1)^2 - 2\right)
\end{equation}
which was previously obtained from a coarse-grained slip-link model.\cite{shanbhag-19}

Combining the free energy of the tube conformation tensor and entanglements results in the total free energy of the entangled melt
\begin{equation}\label{eq:free-energy-dimensional}
	\begin{split}
	F & = \int G_0\bigg[\frac{1}{2}\left(\phi\left(\lambda\right)
		-  \ln \det\tens{A}\right)   \\
		& \qquad + \zeta_Z \nu\left(\ln\nu  - 1\right)\bigg] \diff^3 r,
	\end{split}
\end{equation}
A similar expression for the coarse-grained free energy of a melt with a variable number of entanglements was obtained by \citeauthor{BerisEdwards},\cite{BerisEdwards}  which in our notation reads
\begin{equation}\label{eq:free-energy-beris}
\begin{split}
F & = G_0\int \bigg[\frac{1}{2}\frac{1+\nu Z_{e,eq}}{1 + Z_{e,eq}}(\tr\tens{A} - \ln\det\tens{A})  \\
& \qquad\qquad\quad - \frac{3}{2}\frac{Z_{e,eq}}{1 + Z_{e,eq}}\ln\nu\bigg] \diff^3 r.
\end{split}
\end{equation}
The \citeauthor{BerisEdwards} free energy assumes that the effective plateau modulus depends on the current number of entanglements; the complex form of the pre-factor arises because a completely disentangled melt ($\nu=0$) will still carry some stress.  However, there should be no entanglement-dependent modulus for a conformation tensor defined by \ref{eq:conformation-tensor-definition}, as discussed in Sec.~\ref{sec:im-model}.  The free-energy derivative $\delta F/\delta \nu$ approaches an $\tens{A}$-dependent constant in \eqref{eq:free-energy-beris} as $\nu\to\infty$, whereas $\delta F/\delta\nu\to\infty$ in our model.  This latter behavior is more physical; each entanglement restricts the conformational entropy of the chain, and so it becomes increasingly difficult to add additional entanglements as the number of entanglements increases. 

\subsection{Evolution equations}\label{subsec:evolution-equations}
The evolution equation for the conformation tensor must produce positive entropy in order to satisfy the second law of thermodynamics.  This requirement can be satisfied in the single-bracket formulation of non-equilibrium thermodynamics.\cite{BerisEdwards} which suggests evolution equations of the form 
\begin{widetext}
\begin{subequations}\label{eq:generic}
	\begin{align}
	\begin{split}
	\stackrel{\triangledown}{\tens{A}} & = -\mathbb{M}^{AA}:\frac{\delta F}{\delta \tens{A}} - \tens{M}^{A\nu}\frac{\delta F}{\delta \nu}\label{eq:generic-A}, 
	\end{split}\\
	\begin{split}
		\frac{\D \nu}{\D t} 
		& = -\tens{M}^{\nu A}:\frac{\delta F}{\delta \tens{A}} - M^{\nu\nu} \frac{\delta F}{\delta \nu}, 	\label{eq:generic-nu}
	\end{split}\\
	\begin{split}
		\vec{\sigma} & = \tens{A}\cdot\frac{\delta F}{\delta \tens{A}} + \frac{\delta F}{\delta \tens{A}}\cdot\tens{A} - \nu\tens{I}\frac{\delta F}{\delta \nu}   - \frac{Z_{e,eq}\nu}{Z_{e,eq}\nu + 1}\tens{I}\frac{\delta F}{\delta \tens{A}}:\tens{A},
	\end{split}\label{eq:stress}
	\end{align}
\end{subequations}
\end{widetext}
where $\delta F/\delta \tens{A}$ and $\delta F/\delta \nu$ are  functional derivatives, \new{$\D/\D t=\p/\p t+\vec{v}\cdot\bnabla$ is the material derivative with respect to the velocity $\vec{v}$, and
\begin{equation}
	\stackrel{\triangledown}{\tens{A}} \equiv \frac{\D \tens{A}}{\D t}  - \tens{A}\cdot\bnabla\vec{v} - (\bnabla\vec{v})^\intercal\cdot\tens{A},
\end{equation}
is the upper-convected derivative of the conformation tensor.  A detailed derivation of the stress tensor is given in Appendix~\ref{app:conformation-entanglements-kinematics}. The Appendices present a generalization of the upper-convected derivative to compressible flows which, while irrelevant for most polymer melts, may prove useful for studying (dis)entanglement dynamics in polymer solutions.}

The left-hand sides of \eqref{eq:generic-A} and \eqref{eq:generic-nu} describe the reversible deformation of the melt as derived in Sec.~\ref{sec:theoretical-background}, while the configuration-dependent mobility tensors $\mathbb{M}^{AA}$, $\tens{M}^{\nu A}$ and $M^{\nu\nu}$ (with dimensions of 1/(stress$\cdot$time)) on the right-hand sides describe the dissipative dynamics of the melt. We will see below that the conjugate pair $\tens{M}^{\nu A}$ and $\tens{M}^{A \nu}$  represent the effects of CCR and satisfy Onsager-Casimir symmetry.  We have neglected viscous stresses arising from faster degrees of freedom, which can be accounted for by adding a Newtonian viscous stress to \eqref{eq:stress}.

The functional form of the stress ensures that reversible (elastic) deformations of the melt do not produce entropy.  The mobility tensors $\mathbb{M}^{AA}$, $\tens{M}^{A\nu}$, $\tens{M}^{\nu A}$ and $M^{\nu\nu}$ encode the relaxation processes of the melt, and contain major
\begin{subequations}
	\begin{align}
	M_{ijk\ell}^{AA} & = M_{k\ell ij}^{AA}, \\
	M_{ij}^{\nu A} & = M_{ji}^{A\nu},
	\end{align}
\end{subequations}
and minor
\begin{subequations}
	\begin{gather}
	M_{ijk\ell}^{AA} = M_{jik\ell}^{AA} = M_{ij\ell k}^{AA}, \\
	M_{ij}^{\nu A} = M_{ji}^{\nu A}
	\end{gather}
\end{subequations}
symmetries consistent with the Onsager-Casimir reciprocal relations.\cite{onsager-31a,onsager-31b,casimir-45}  \new{These symmetries are embedded in the dissipation bracket of the single generator formalism.\cite{BerisEdwards}}

The stress tensor \eqref{eq:stress} can be computed using the free energy \eqref{eq:free-energy-dimensional}, yielding
\begin{equation}\label{eq:stress-melt}
	\begin{split}
		\vec{\sigma} & = G_0\bigg(\f(\lambda) \tens{A} - \tens{I} - \zeta_Z\nu \ln\nu \vec{I} \\
			& \qquad\qquad - \frac{3}{2}\frac{Z_{e,eq}\nu(\f(\lambda)\lambda^2 - 1)}{Z_{e,eq}\nu + 1}\tens{I} \bigg),
	\end{split}
\end{equation}
where 
\begin{equation}
\f(\lambda) = \frac{\p \phi(\lambda)}{\p \tr\tens{A}}
\end{equation}
is the spring force, which becomes
\begin{equation}
\f(\lambda) = 1 + \frac{2}{3}\frac{\lambda^2 - 1}{\lambda_\textrm{max}^2 - \lambda^2}
\end{equation}
for the Cohen approximation shown in \eqref{eq:phi}.   The limit $\lambda_\textrm{max}\to\infty$ corresponds to an infinitely extensible chain, yielding $\f=1$.  The first two terms in parenthesis in \eqref{eq:stress-melt} comprise the usual stress for a polymer melt or solution with finite extensibility.  The third and fourth terms both arise from the entanglements, and contribute to the pressure.

The dissipative dynamics of the polymer stretch, orientation, and entanglements all arise from polymer drag, suggesting connections between the relaxation dynamics of these quantities.  Following \citet{im-14}, we assume entanglements are removed via convective constraint release as the chain stretches.  It follows that $\tens{M}^{\nu A}$, which encodes flow-induced disentanglement of the melt, and $\mathbb{M}^{AA}$, which encodes relaxation of the polymer stretch and orientation, should contain similar physics.  Furthermore, the rate of disentanglement should tend to zero as the number of entanglements per unit length on the stretched chain $\nu/\lambda$ becomes small.  It is therefore reasonable to assume that $\tens{M}^{\nu A}\sim (\nu/\lambda)\mathbb{M}^{AA}:\p\lambda/\p\tens{A}$, where the contraction with the partial derivative extracts the mobility in the direction the tube is stretching.  The most general forms of the mobility tensors consistent with these physical arguments are
\begin{subequations}
	\begin{align}
	\tens{M}^{\nu A} & = \frac{\beta\nu}{\lambda} \frac{\p\lambda}{\p \tens{A}}:\mathbb{M}^{AA} \label{eq:MAnu},  \\
	\begin{split}
	M^{\nu\nu} & = \left(\frac{\beta\nu}{\lambda} \right)^2  \frac{\p\lambda}{\p \tens{A}}:\mathbb{M}^{AA}:\frac{\p\lambda}{\p \tens{A}} \\
		& \quad + \frac{1}{\zeta_Z G_0 \tau_\nu},
	\end{split} \label{eq:Mnunu}
	\end{align}
\end{subequations}
where $\mathbb{M}^{AA}$ is the same as in \eqref{eq:generic-A}, $\beta$ is a constant controlling the rate of disentanglement, and $\tau_\nu$ is the, as yet undetermined, re-entanglement time.

The terms proportional to $\mathbb{M}^{AA}$ in $M^{\nu\nu}$ ensure consistency with the second law of thermodynamics.  The entropy production for scalar entanglements and a tensorial tube conformation tensor is given by\cite{BerisEdwards,ps-04}
\begin{equation}\label{eq:entropy-production}
\begin{split}
T\dot s & = -\frac{\delta F}{\delta \tens{A}}:\stackrel{\triangledown}{\tens{A}}
- \frac{\delta F}{\delta \nu}\frac{\D\nu}{\D t}\\
& = 
\left(\begin{matrix}
\frac{\delta F}{\delta \tens{A}} &
\frac{\delta F}{\delta \nu}
\end{matrix}\right)\left(\begin{matrix}
\mathbb{M}^{AA} &   \tens{M}^{\nu A} \\
\tens{M}^{\nu A} & M^{\nu\nu}
\end{matrix}\right)
\left(\begin{matrix}
\frac{\delta F}{\delta \tens{A}}\\[5truept]
\frac{\delta F}{\delta \nu}
\end{matrix}\right).
\end{split}
\end{equation} 
Substituting \eqref{eq:MAnu} and \eqref{eq:Mnunu} into \eqref{eq:entropy-production} yields 
\begin{widetext}
	\begin{equation}
	T\dot s = \left(\frac{\delta F}{\delta \tens{A}} + \frac{\beta \nu}{\lambda}\frac{\p \lambda}{\p \tens{A}}\frac{\delta F}{\delta \nu}\right) :
	\mathbb{M}^{AA} : \left(\frac{\delta F}{\delta \tens{A}} +\frac{\beta \nu}{\lambda}\frac{\p \lambda}{\p \tens{A}}\frac{\delta F}{\delta \nu}\right) + \frac{\delta F}{\delta \nu}\frac{1}{\zeta_Z G_0 \tau_\nu}\frac{\delta F}{\delta \nu},
	\end{equation}
\end{widetext}
which is non-negative if $\mathbb{M}^{AA}$ is positive definite and~{$\tau_\nu>0$}.

One thus obtains the following evolution equations for the conformation tensor and normalized entanglement number:
\begin{subequations}\label{eq:generic-Mob}
	\begin{align}
	\stackrel{\triangledown}{\tens{A}}  &= -\mathbb{M}^{AA}:\left(\frac{\delta F}{\delta \tens{A}} 
	+ \frac{\beta\nu}{\lambda} \frac{\p\lambda}{\p \tens{A}}\frac{\delta F}{\delta \nu}\right), \label{eq:gen-M-A}\\
	\begin{split}
	\frac{\D\nu}{\D t}  &=
	- \frac{\beta\nu}{\lambda} \frac{\p\lambda}{\p \tens{A}}:\mathbb{M}^{AA}:\left(\frac{\delta F}{\delta \tens{A}} 
	+ \frac{\beta\nu}{\lambda} \frac{\p\lambda}{\p \tens{A}}\frac{\delta F}{\delta \nu}\right) \\
		 &\quad - \frac{\ln\nu}{\tau_\nu}.
	\end{split}\label{eq:gen-M-nu}
	\end{align}
\end{subequations}
The entanglement evolution can be re-expressed as
\begin{equation}\label{eq:generaic-im}
	\frac{\D\nu}{\D t}   = \frac{\beta\nu}{\lambda}\frac{\p \lambda}{\p \tens{A}}:\stackrel{\triangledown}{\tens{A}} - \frac{\ln\nu}{\tau_\nu},
\end{equation}
from which we see that the physical and thermodynamic arguments used to obtain \eqref{eq:MAnu} and \eqref{eq:Mnunu} exactly reproduce the I-M disentanglement mechanism derived in \eqref{eq:im-upper-convected}.  The second term on the right-hand side accounts for re-entanglement.  Furthermore, it prevents the number of entanglements from becoming negative because $\ln\nu$ becomes infinite as $\nu\to0$.  Near equilibrium, 
\begin{equation}
	\frac{\ln\nu}{\tau_\nu} \simeq \frac{\nu-1}{\tau_\nu},
\end{equation}
which is similar to the $(\nu-1)/\tau_{d,eq}$ term in the I-M model.  However, as will be shown in Sec.~\ref{subsec:tau-nu}, \new{simulations suggest that} $\tau_\nu\ll\tau_{d,eq}$, \new{implying that} re-entanglement occurs faster than predicted by \citet{im-14}.

\subsection{Onsager-Casimir reciprocity}\label{subsec:onsager-casimir}
The formulation of our disentanglement model begins with \eqref{eq:generic-nu}, which states that relaxation of a melt is driven by changes in the free energy. This appears fundamentally different from the Ianniruberto-Model, which in \eqref{eq:im-model} predicts that the rate of disentanglement is proportional to the non-affine stretch rate of the melt. However, we showed in \eqref{eq:generaic-im} that the Onsager-Casimir relations imply that these two forms are equivalent when one makes physically motivated assumptions about the form of $\tens{M}^{\nu A}$ and $M^{\nu\nu}$.  This equivalency is a generic features of all constitutive equations consistent with the Onsager-Casimir reciprocity because one can always use \eqref{eq:generic-A} to express $\delta F/\delta\tens{A}$ in terms of the upper-convected derivative.

\new{Similarly, the Onsager-Casimir reciprocal relations dictate that the influence of entanglement on the conformation tensor can be equivalently modeled through either the free-energy derivative $\delta F/\delta\nu$ or through the disentanglement rate $D\nu/Dt.$ Consequently, disentanglement will reduce the conformation tensor (and hence the stress) when $\beta>0$ as expected on physical grounds (by expressing $\delta F/\delta\nu$ in Eq.~(\ref{eq:gen-M-A}) in terms of $D\nu/Dt$).  These dynamics can be made explicit by re-arranging \eqref{eq:gen-M-A} and \eqref{eq:gen-M-nu} to obtain
 \begin{equation}\label{eq:pdo}
    \begin{split}
			\stackrel{\triangledown}{\tens{A}}  & = -\left(\mathbb{M}^{AA} - \frac{\mathbb{M}^{AA}:\frac{\p\lambda}{\p\tens{A}}\frac{\p\lambda}{\p\tens{A}}:\mathbb{M}^{AA}}{\frac{\p\lambda}{\p\tens{A}}:\mathbb{M}^{AA}:\frac{\p\lambda}{\p\tens{A}}}\right):\frac{\delta F}{\delta \tens{A}} \\
			& \quad + \frac{\lambda}{\beta\nu}\frac{\mathbb{M}^{AA}:\frac{\p\lambda}{\p\tens{A}}}{\frac{\p\lambda}{\p\tens{A}}:\mathbb{M}^{AA}:\frac{\p\lambda}{\p\tens{A}}}\left(\frac{\D\nu}{\D t} + \frac{\ln\nu}{\tau_\nu}\right),
    \end{split}
	\end{equation}
The influence of the thermodynamic driving force $\delta F/\delta\nu$ in the conformational dynamics (or as manifested by the dynamics $\D\nu/D t$}) is absent in the I-M model. The I-M model is hence inconsistent with Onsager-Casimir reciprocity.  On the other hand, \new{\citet{hhhr-15} argued that fast equilibrium of Kuhn segments within the tube implies that the stress remains constant during re-entanglement, which (in their model) required adding an additional term to the conformation dynamics when $\D\nu/\D t>0$.  However, Onsager-Casimir reciprocity requires that the conformational and entanglement dynamics couple for all values of $\D\nu/\D t$.  In our model, $\D\nu/\D t+\ln\nu/\tau_\nu$ vanishes when no disentanglement occurs, and the asymmetry between disentanglement and re-entanglement arises naturally from the chosen form of $M^{\nu\nu}$ and the Onsager-Casimir reciprocity.}

\subsection{Mobility tensors}\label{subsec:mobility}
In Appendix~\ref{app:green-kubo} we derive a physically motivated expression for the mobility tensor based on the Green-Kubo relations:
\begin{equation}\label{eq:MAA-general}
\begin{split}M_{ijk\ell}^{AA} & = 
\frac{1}{2}\left(M^\textrm{rep}_{ik}A_{j\ell} + M^\textrm{rep}_{i\ell}A_{jk} \right. \\
& \quad \left. + A_{ik}M^\textrm{rep}_{j\ell} + A_{i\ell}M^\textrm{rep}_{jk}\right)
+ \frac{4}{\lambda^2}M^\textrm{ret}A_{ij}A_{k\ell},
\end{split}
\end{equation}
where the mobility $\tens{M}^\textrm{rep}$ arises from reptation and the mobility $M^\textrm{ret}$ describes retraction and contour-length fluctuations.  The conformation mobility tensor is positive definite, and hence consistent with the second law of thermodynamics, provided $\tens{A}$ and $\tens{M}^\textrm{rep}$ are positive definite and
\begin{equation}
	\lambda^2 + 2 M^\textrm{ret}\tr[\tens{A}\cdot(\tens{M}^\textrm{rep})^{-1}]>0.
\end{equation}	
This later condition is automatically satisfied provided $M^\textrm{ret}>0$.  From \eqref{eq:MAnu} and \eqref{eq:Mnunu}, the corresponding cross-coupling and entanglement mobilities are
\begin{subequations}
	\begin{align}
		\tens{M}^{\nu A} & = \frac{\beta\nu}{3\lambda^2}\left(\tens{A}\cdot\tens{M}^\textrm{rep}
		+ 6M^\textrm{ret}\tens{A}\right),  \\
		\begin{split}
			M^{\nu\nu} & = \left(\frac{\beta\nu}{3\lambda^2} \right)^2 \bigg(\frac{1}{2}\tr\left(\tens{A}\cdot\tens{M}^\textrm{rep}\right)
			+ 9\lambda^2 M^\textrm{ret}\bigg)\\
				& \quad + \frac{1}{\zeta_Z G_0 \tau_\nu},
		\end{split}
	\end{align}
\end{subequations}

\new{The formulation thus far is quite general. We next} take inspiration from the Rolie-Poly \new{model} to obtain a specific mobility tensor. The Rolie-Poly \new{model}\cite{lg-03} incorporates the three primary relaxation mechanisms \new{(CCR, retraction, and reptation)} into the  kinetic equation for the conformation tensor $\tens{A}$:
\begin{equation}\label{eq:rolie-poly}
	\stackrel{\triangledown}{\tens{A}}_{\scriptscriptstyle \textrm{RP}}=
	-\frac{1}{\tau_d(\lambda)}(\tens{A}_{\scriptscriptstyle \textrm{RP}} - \tens{I}) - \frac{2}{\tau_R}\left(1 - \frac{1}{\lambda} \right) \tens{A}_{\scriptscriptstyle \textrm{RP}},
\end{equation}
where $\vec{\sigma} = G_0(\tens{A} - \tens{I})$ and $\lambda=\sqrt{\tr\tens{A}/3}$.  The first term encodes both reptation and CCR, while the second term encodes retraction of the tube.  The non-equilibrium reptation relaxation rate is
\begin{equation}
\frac{1}{\tau_d(\lambda)} = \frac{1}{\tau_{d,eq}} + 2\frac{\beta}{\tau_R}(\lambda - 1)\lambda^{2\delta - 1},
\end{equation}
where the first term is the equilibrium reptation rate and the second term accounts for faster relaxation due to CCR.  The value  $\delta=-0.5$ was proposed  in the original Rolie-Poly derivation\cite{lg-03} and is employed throughout this work.

The Rolie-Poly equation is consistent with an isotropic reptation mobility tensor $\tens{M}^{\textrm{rep}}$, which leads to a model that predicts zero second normal stress differences.  Instead, we employ an anisotropic \citet{giesekus-82} mobility
\begin{equation}\label{eq:Giesekus}
	\tens{M}^\textrm{rep} = \frac{1}{G_0\tau_d(\lambda)}\left((1 - \alpha)\tens{I} + \alpha\tens{A}\right),
\end{equation}
where $\alpha$ parameterizes the anisotropy.  The retraction mobility consistent with the Rolie-Poly equation is
\begin{equation}\label{eq:M-ret}
	M^\textrm{ret} = \frac{1}{3G_0\tau_R}\frac{\lambda}{\lambda + 1},
\end{equation}
which increases with increasing stretch, which may reflect the more ordered segments in the stretched state.
\begin{widetext}
The form of $\mathbb{M}^{AA}$ consistent with the Rolie-Poly model is then
\begin{equation}\label{eq:MAA}
\begin{split}
M_{ijk\ell}^{AA} & = 
\frac{1 - \alpha}{2G_0\tau_d(\lambda)}\Big(\delta_{ik}A_{j\ell} + \delta_{i\ell}A_{jk} + A_{ik}\delta_{j\ell} + A_{i\ell}\delta_{jk}\Big) 
	+ \frac{\alpha}{G_0\tau_d(\lambda)}\Big(A_{ik}A_{j\ell} + A_{i\ell}A_{jk}\Big)  \\
	& \quad + \frac{4}{3G_0\tau_R} \frac{1}{\lambda^2 + \lambda} A_{ij}A_{k\ell}.
\end{split}
\end{equation}
The reptation mobility in \eqref{eq:MAA} is very similar to the reptation mobility of \citeauthor{stm-16},\cite{stm-16}, which arises from an isotropic diffusive process; but the retraction mobility differs. \citet{stm-16} constructed a retraction mobility similar in form to the reptation mobility, while we propose an anisotropic  retraction mobility that only allows retraction  within the tube. From this mobility we can obtain $\tens{M}^{\nu A}$ and $M^{\nu\nu}$:
\begin{subequations}
    \begin{align}
        \tens{M}^{\nu A} &= \frac{\beta\nu}{G_03\lambda^2}
			\left(\frac{\tens{I}}{\tau_d(\lambda)} + \alpha\frac{\tens{A} - \tens{I}}{\tau_d(\lambda)} 
				+ \frac{2\lambda \tens{I}}{\tau_R(\lambda + 1)}\right)\cdot\tens{A}\\
        M^{\nu\nu} &= \frac{1}{2G_0}\left(\frac{\beta\nu}{3\lambda^2}\right)^2
			\left(\frac{\tens{I}}{\tau_d(\lambda)} + \alpha\frac{\tens{A} - \tens{I}}{\tau_d(\lambda)} 
				+ \frac{2\lambda \tens{I}}{\tau_R(\lambda + 1)}\right):\tens{A} + \frac{1}{\zeta_Z G_0 \tau_\nu}.
    \end{align}
\end{subequations}

\color{black}\subsection{Governing equations}\label{subsec:governing-equation}
\color{black}

\new{Putting all of the ingredients above together, we obtain the following equations of motion, which is one of the primary results of this work:}
\begin{subequations}\label{eq:governing-equations}
	\begin{align}
		\begin{split}
			\stackrel{\triangledown}{\tens{A}}  & = 
				\new{- \left(\frac{\tens{I}}{\tau_d(\lambda)} + \alpha\frac{\tens{A}-\tens{I}}{\tau_d(\lambda)}\right)\cdot(\f(\lambda)\tens{A} - \tens{I}) 
				- \frac{2}{\tau_R} \frac{\f(\lambda) \lambda^2 - 1}{\lambda^2 + \lambda}\tens{A} - \frac{\zeta_Z\beta\nu}{3\lambda^2}
			\left(\frac{\tens{A}}{\tau_d(\lambda)} + \alpha\frac{\tens{A^2} - \tens{A}}{\tau_d(\lambda)} 
				+ \frac{2\lambda \tens{A}}{\tau_R(\lambda + 1)}\right)\ln\nu,}
	\end{split}\label{eq:governing-equation-A}\\
	\begin{split}
		\frac{\D \nu}{\D t} & = -\frac{\beta\nu}{3\lambda^2}\left( \tens{A}:\bnabla \vec{v} - \frac{1}{2}\frac{\D\tr\tens{A}}{\D t}\right) 	- \frac{\ln\nu}{\tau_\nu},		
	\end{split}\label{eq:governing-equation-nu} \\
	\vec{\sigma} & = G_0\left(\f(\lambda)\tens{A} - \tens{I}  - \frac{3}{2}\frac{Z_{e,eq}\nu(\f(\lambda)\lambda^2 - 1)}{Z_{e,eq}\nu + 1}\tens{I} - \zeta_Z\nu\ln\nu \tens{I}\right).
	\end{align}
\end{subequations}
Here, we have used the identification in \eqref{eq:generaic-im} to write the entanglement dynamics in the simpler form that does not explicitly use $\tens{M}^{\nu A}$ or $M^{\nu\nu}$.  An alternate formulation of the governing equations in terms of stretch and orientation is given in Appendix~\ref{app:stretch-orientation}.

In the absence of FENE effects $(\f=1)$ and second normal stress differences ($\alpha=0$), these equations become
\begin{subequations}\label{eq:governing-equations-no-FENE}
	\begin{align}
	 \stackrel{\triangledown}{\tens{A}}  & = 
	- \frac{1}{\tau_d(\lambda)}(\tens{A} - \tens{I}) 
	- \frac{2}{\tau_R} \left(1 - \frac{1}{\lambda} \right)\tens{A}
	- \frac{\zeta_Z\beta\nu}{3\lambda^2}
	\left(\frac{1}{\tau_d(\lambda)}
	+ \frac{2\lambda}{\tau_R(\lambda + 1)}\right)\tens{A}\ln\nu, \label{eq:governing-equation-A-no-FENE}\\
	\frac{\D \nu}{\D t}  & = -\frac{\beta\nu}{3\lambda^2}\left( \tens{A}:\bnabla \vec{v} - \frac{1}{2}\frac{\D\tr\tens{A}}{\D t}\right)  - \frac{\ln\nu}{\tau_\nu}. \label{eq:governing-equation-nu-no-FENE} \\
	\vec{\sigma} & = G_0\left(\tens{A} - \tens{I} - \frac{3}{2}\frac{Z_{e,eq}\nu(\lambda^2 - 1)}{Z_{e,eq}\nu + 1}\tens{I}  - \zeta_Z\nu\ln\nu \tens{I}\right).
	\end{align}
\end{subequations}
.
\end{widetext}
This formulation is equivalent to the Rolie-Poly equation \eqref{eq:rolie-poly}, with additional contributions from the entanglement kinetics.  

The first two terms in \eqref{eq:governing-equation-A} \new{or \eqref{eq:governing-equation-A-no-FENE}} recover the Rolie-Poly equation. The final term in \eqref{eq:governing-equation-A}  arises from a non-equilibrium ($\nu\neq1$) number of entanglements. A disentangled melt ($\ln\nu<0$) drives the conformation tensor \textit{away} from equilibrium. Put simply, \new{stretch relaxation} drives disentanglement and disentanglement drives further stretch.  This behavior is a consequence of the Onsager-Casimir reciprocal relations, and must be present in any thermodynamically consistent consistent theory of disentanglement to ensure microscopic reversibility. It will be shown in Sec.~\ref{sec:results} that this leads to a small increase in the stress at high shear rates.

For thermodynamic consistency $\mathbb{M}^{AA}$ must be positive definite.  This is satisfied if $\tau_d(\lambda)>0$, $0\leq\alpha<1$, and $\tens{A}$ is positive definite.  A proof due to \citet{hulsen-90} shows that a positive definite $\tens{A}$ will remain positive definite if $\tau_d(\lambda)>0$.  In most flows $\lambda>1$ and a positive reptation function is automatically satisfied provided $\beta>0$.  The stretch can decrease very slightly below one in decelerating flows,\cite{rtmt-17} however,  the reptation function will remain positive for such small compression.  Our model is thus thermodynamically consistent.

\subsection{The re-entanglement time}\label{subsec:tau-nu}
\begin{figure}
	\centering
	\includegraphics[width=0.45\textwidth]{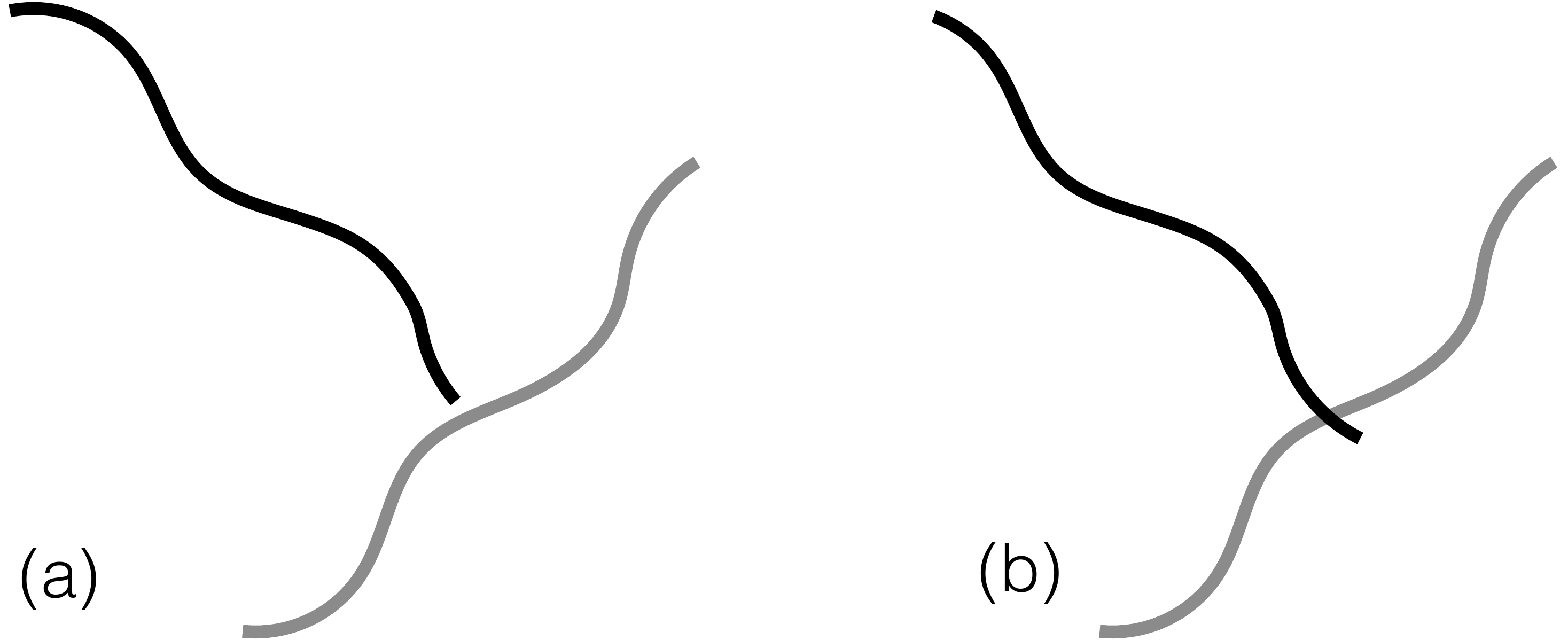}
	\caption{Representation of entanglement formation.  Two initially unentangled chains (a) come together to form a new entanglement (b). An entanglement formed at the end of one chain is an entanglement formed in the interior of a second chain.}\label{fig:entanglement-reptation}
\end{figure}

\begin{figure}
	\centering
	\begin{minipage}[b]{.02\textwidth}
		(a) 
		\vspace{120pt}
	\end{minipage}
	\includegraphics[width=0.43\textwidth]{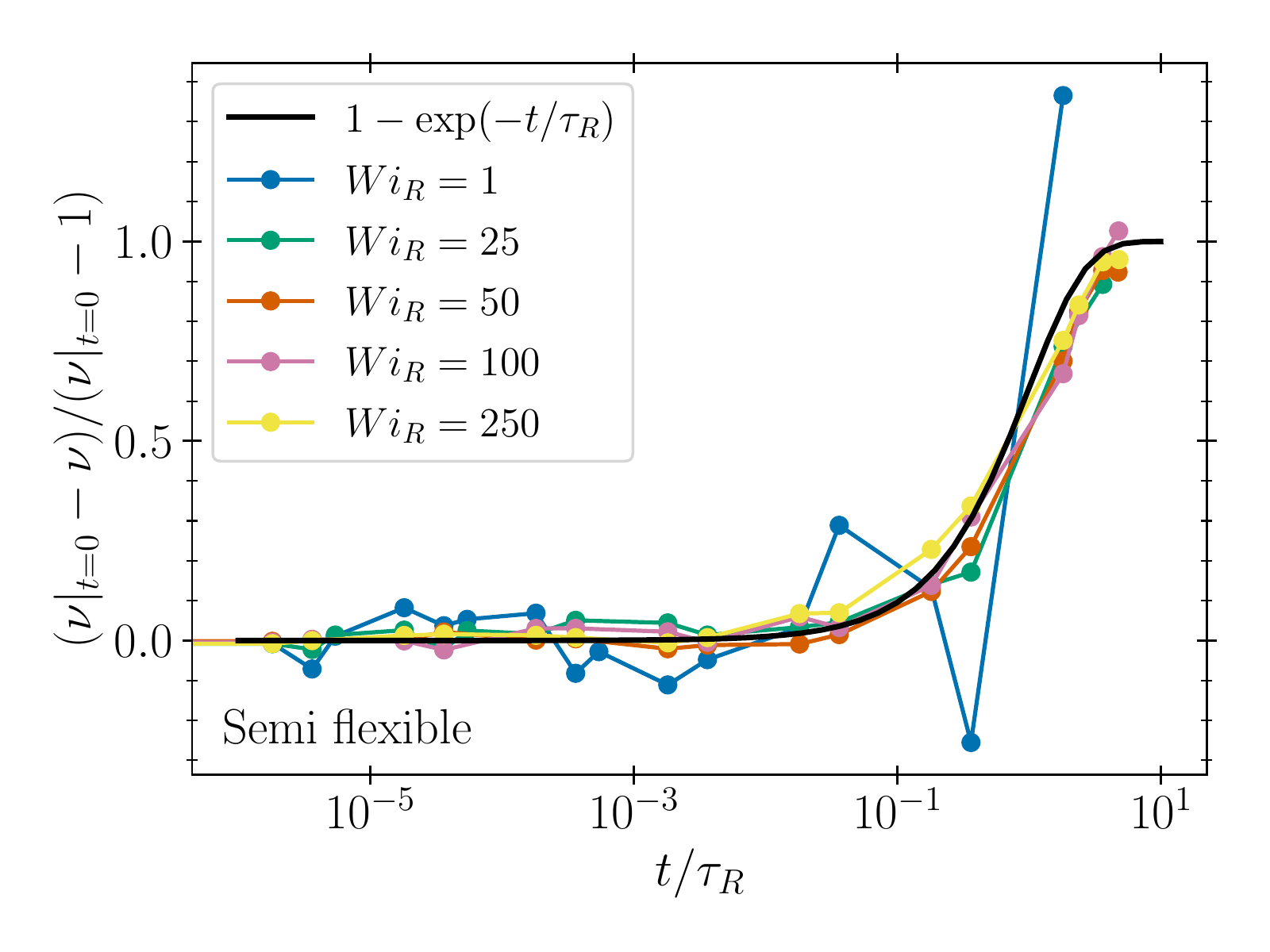}\\
	\begin{minipage}[b]{.02\textwidth}
		(b) 
		\vspace{120pt}
	\end{minipage}
	\includegraphics[width=0.43\textwidth]{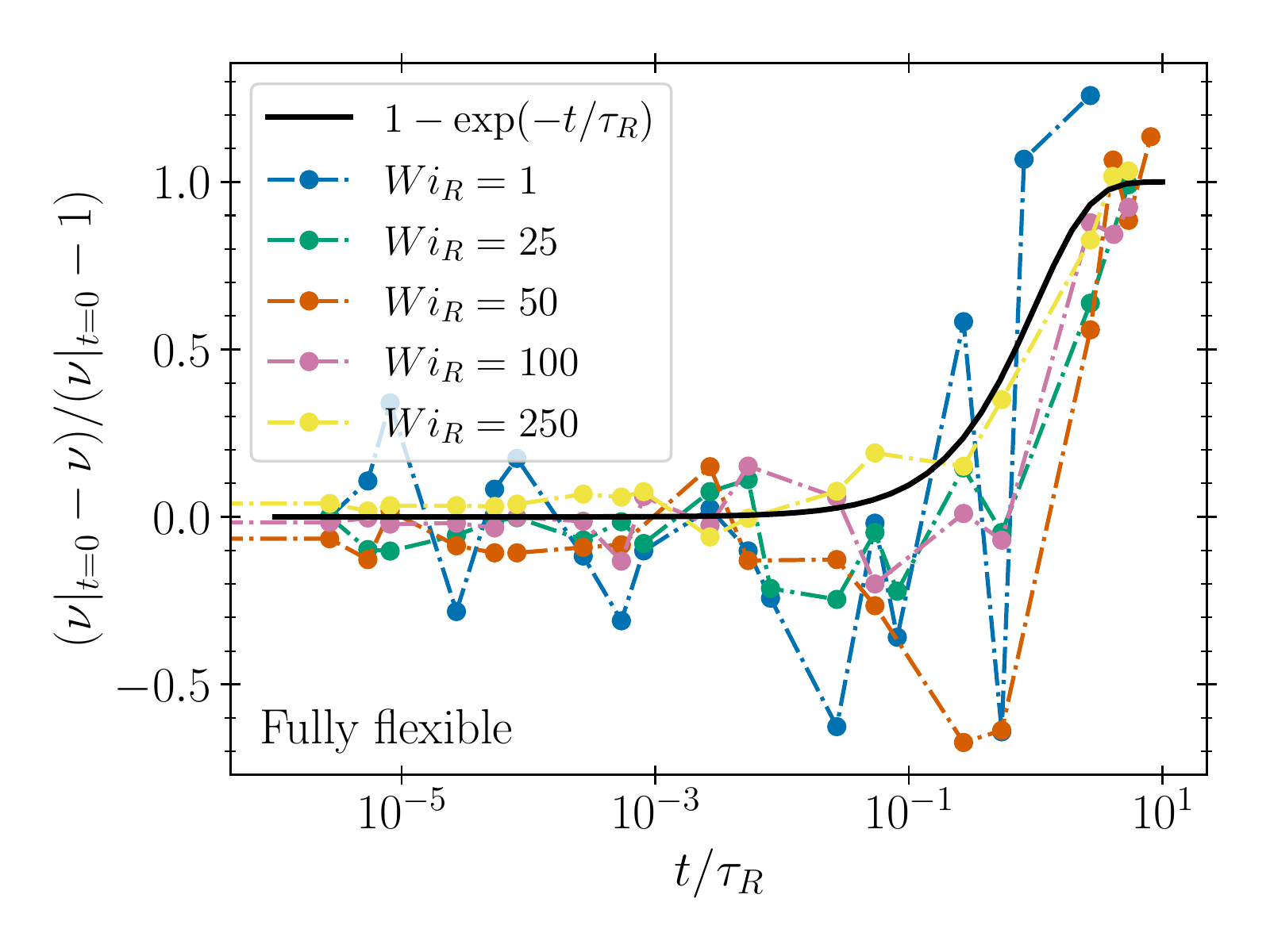}
	\caption{Recovery of entanglement after cessation of steady shear in molecular dynamics simulations\cite{Marco_Thesis} of Kremer-Grest melts as a function of time.  (a) semi-flexible melt with $Z_{e,eq}\simeq35$ (b) fully flexible melt with $Z_{e,eq}\simeq11$.}\label{fig:tau-nu-timescales}
\end{figure}

Prior studies\cite{im-14,hhhr-15,mbp-15,mmp-18} assumed that entanglements in the melt recover via slow reptation.  In the original tube model,\cite{DoiEdwards_Book} entanglements are equally spaced along the tube, and reform at chain ends via reptation.  This mechanism is consistent with a re-entanglement time $\tau_\nu=\tau_{d,eq}$ as in the I-M model.  However, this simple picture neglects the formation of entanglements in the interior of the chain due to the reptation of a second chain. Scaling theories consistent with experimental data suggest that entanglements are binary interactions between chains.\cite{milner-20}  As illustrated in figure~\ref{fig:entanglement-reptation}, the rates of  tube renewal at the chain ends and formation of a new entanglement in the chain center must be identical if entanglements represent binary interactions.  Indeed, \citet{ks-09} used the equality of these rates as a self-consistency check in their slip-link dynamics.  This equality implies that  $\tau_\nu=\tau_{d,eq}/2$ if re-entanglement is governed by reptation, as proposed in prior studies.\cite{mbp-15,mmp-18}

However, these simple arguments contradict molecular dynamics simulations which imply that entanglements recover on the Rouse time.\cite{ohr-19,Marco_Thesis,bsek-22}  This faster re-entanglement may arise as follows.  A stochastic increase in chain length on the Rouse time $\tau_R$ can increase the length of the dangling ends of the tube, leading to formation of new entanglements at the chain ends over a timescale $\tau_e$.  At thermodynamic equilibrium, the rates of creation and destruction of entanglements via contour length fluctuations must be identical, and contour length fluctuations simply decrease the effective reptation time.\cite{lm-02}  In a disentangled melt, this mechanism will lead to creation of new entanglements at \new{both the ends and interior of chains as illustrated in figure~\ref{fig:entanglement-reptation}}.  The re-entanglement time $\tau_\nu$ will thus be proportional to the Rouse time under this proposed mechanism. We make the parsimonious choice that these two time scale are identical, i.e.
\begin{equation}
    \tau_\nu = \tau_R.
\end{equation}

This argument does not imply that the conformation tensor, and hence the stress, will relax on the Rouse time.  Consider the re-entanglement of a melt following cessation of a shear flow.  The melt will mostly re-entangle over a Rouse time due to formation of entanglement near the chain ends.  However, relaxation of the conformation tensor (and hence stress) requires relaxation of the interior strands, which can only occur via reptation.  Thus, the re-entangled melt is still in an anisotropic state under our proposed mechanism, and re-entanglement can occur without fully relaxing the stress.  This phenomena has been observed in molecular dynamics simulations.\cite{ohr-19} 

\new{Our coarse-grained description cannot  distinguish whether or not a given entanglement fraction $\nu$ is distributed uniformly or non-uniformly along the tube coordinate $s$. During re-entanglement after cessation of flow (for example) new entanglements will arise at the end of a chain (and correspondingly at other positions in entangling partners). Entanglements entering from an end should move along the tube controlled by a combination of Rouse and reptation dynamics, in addition to possible motions such as interchain slithering and winding. However, these dynamics are not captured in this model. The simulations to which we compare our results also have not resolved this level of detail (although in principle such an analysis could be performed). A more detailed description of entanglement dynamics is likely required to explain recent molecular dynamics simulations of entanglement recovery following flow cessation, where $\nu$ recovers to almost its equilibrium value on the Rouse time, followed by a final slow recovery on the reptation time.\cite{bsek-22}}

To test this mechanism we examine the re-entanglement observed in simulations\cite{Marco_Thesis} of two Kremer-Grest\cite{kg-90} melts as a function of the Rouse Weissenberg number $\textrm{Wi}_R=\dot\gamma\tau_R$: a semi-flexible melt with $Z_{e,eq}\simeq35$, and a fully flexible melt with $Z_{e,eq}\simeq11$.  Re-entanglement of both melts following cessation of steady shear flow is shown in figure~\ref{fig:tau-nu-timescales}.  We observe that entanglements approximately recover exponentially on the Rouse time.  The exponential recovery fits the data well for both melts, which suggests that $\tau_\nu\sim\tau_R$ rather than $\tau_\nu\sim\tau_{d,eq}$ as assumed in prior studies.  The agreement for the semi-flexible melt is more apparent at high $\textrm{Wi}_R$ because of entanglement fluctuations at low $\textrm{Wi}_R$.  \new{Exponential recovery of entanglements on the Rouse time has also been observed in united-atom polyethylene simulations.\cite{bsek-22}}

\section{Analytical solutions}\label{sec:analytical}
In this section, we derive analytical solutions of the model presented in Sec.~\ref{sec:thermodynamic}.  In Sec.~\ref{subsec:linear-response} and Sec.~\ref{subsec:weakly-nonlinear} asymptotic solutions are obtained in the limit of weak flows.  Finally, we derive an exact solution for the steady-state entanglement density in Sec.~\ref{subsec:nu-analytical}

\subsection{Linear response}\label{subsec:linear-response}

To obtain asymptotic solutions of our model in the limit of slow flows we expand \eqref{eq:governing-equation-A} and \eqref{eq:governing-equation-nu} about equilibrium ($\tens{A}=\tens{I},\nu=1$).  It is convenient to separate the conformation tensor into its deviatoric and isotropic parts
\begin{equation}
	\tens{A} = \tens{A}^\textrm{dev} + \frac{\tr\tens{A}}{3}\tens{I}.
\end{equation}
Expanding $\tens{A}$ about $\tens{I}$ is equivalent to expanding in $\tens{A}^\textrm{dev}$ and $\tr\tens{A}-3=3(\lambda^2-1)$.  We find
\begin{widetext}
		\begin{subequations}\label{eq:governg-equation-linear}
			\begin{align}
			\begin{split}
			\frac{\D\tens{A}^\textrm{dev}}{\D t}   & = 2\tens{D} - \frac{1}{\tau_{d,eq}}\tens{A}^\textrm{dev}, 
			\end{split}\label{eq:governing-equation-A-linear}\\
		\frac{\D}{\D t}\left(\begin{matrix}
			\new{\lambda^2 -1} \\
			\new{\nu - 1}
		\end{matrix}\right)   &= - \frac{1}{3}\left(\frac{1}{\tau_{d,eq}} + \frac{1}{\tau_R}\right)\left(\begin{matrix} 3
			\frac{\f^\prime(1) +2}{2}
				&  \zeta_Z\beta \\
				3\beta\frac{\f^\prime(1) +2}{4}
				&  \frac{\zeta_Z\beta^2}{2}\end{matrix}\right)\left(\begin{matrix}
		\lambda^2 - 1\\
			\nu - 1
		\end{matrix}\right) - \frac{1}{\tau_\nu}\left(\begin{matrix}
			0 \\
			\nu-1
		\end{matrix}\right)\label{eq:governing-equations-stretch-linear}\\
			\vec{\sigma} & = G_0\tens{A}^\textrm{dev} + G_0\left(\frac{\f^\prime(1) + 2}{2}(\lambda^2 - 1) - \zeta_Z(\nu - 1)\right)\tens{I}  - \frac{3}{2}\frac{Z_{e,eq}}{Z_{e,eq} + 1}\frac{\f^\prime(1) + 2}{2}(\lambda^2 - 1) \tens{I},\label{eq:stress-linear}
			\end{align}
		\end{subequations}
\end{widetext}
where $\tens{D}$ is the symmetric part of the velocity gradient. Hence, \new{to lowest order the stretch and entanglement dynamics decouple from the conformational dynamics, flow, and stress} (note that $f^\prime(1)=0$ in the absence of FENE effects).

The linear rheological response is obtained by substituting a spatially uniform, time-dependent velocity field 
\begin{equation}
(\bnabla\vec{v})^\intercal = \vec{\kappa}(t),
\end{equation}
into \eqref{eq:governg-equation-linear}, leading to
\begin{equation}
\vec{\sigma} = 2 \int_{-\infty}^tG(t-t^\prime)\tens{D}(t^\prime) \diff t,
\end{equation}
where the dynamic modulus is of the Maxwell form:
\begin{equation}\label{eq:linear-modulus}
	G(t) = G_0\exp(-t/\tau_{d,eq}).
\end{equation}

\subsection{Weakly nonlinear response}\label{subsec:weakly-nonlinear}
We next compute the conformation tensor and entanglements to quadratic order in the velocity gradient for steady-state homogeneous flows.  We expand $\tens{A}^\textrm{dev}$, $\lambda^2$, and $\nu-1$ in powers of $\textrm{Wi}=\dot\gamma \tau_{d,eq}\ll1$ and solve \eqref{eq:governing-equations} perturbatively to $O(\Wi^2)$ for constant, homogeneous velocity gradient $(\bnabla\vec{v})^\intercal=\vec{\kappa}$.  The $O(Wi)$ perturbations to $\lambda^2$ and $\nu$ vanish because there is no coupling between flow and stretch and orientation to linear order.  The solutions, in dimensional variables, are
\begin{widetext}

\begin{subequations}\label{eq:weakly-nonlinear}
	\begin{align}
	\begin{split}
	\tens{A} & = \tens{I} + 2\tau_{d,eq}\tens{D}- 4\alpha\tau_{d,eq}^2\left( \tens{D}^2 - \tens{I}\frac{\tr\tens{D}^2}{3} \right) + 2\tau_{d,eq}^2 \left(\vec{\kappa}^\intercal\cdot\tens{D} + 
	\tens{D}\cdot\vec{\kappa}
	 - \frac{2}{3}\tens{I}\tr\tens{D}^2\right)  \\
	& \quad + \frac{8\tau_{d,eq}^2}{3(\f^\prime(1) + 2)}\left(\frac{(1-\alpha)\tau_R}{\tau_{d,eq} + \tau_R} + \frac{\zeta_Z\beta^2\tau_\nu}{6\tau_{d,eq}} \right)\tens{I}\tr\tens{D}^2, \\
	\end{split} \label{eq:A-weakly-nonlinear} \\
	\nu & = 1 - \frac{2\beta}{3}\tau_{d,eq}\tau_\nu\tr\tens{D}^2 \label{eq:nu-weakly-nonlinear}\\
	\lambda^2 &= 1 + \frac{8\tau_{d,eq}^2}{3(\f^\prime(1) + 2)}\left(\frac{(1-\alpha)\tau_R}{\tau_{d,eq} + \tau_R} + \frac{\zeta_Z\beta^2\tau_\nu}{6\tau_{d,eq}} \right) \tr\tens{D}^2.
	\end{align}
\end{subequations}
The resultant stress tensor is
\begin{align}
		\vec{\sigma}   & = 2G_0\tau_{d,eq}\tens{D} - 4\alpha\tau_{d,eq}^2\tens{D}^2 + 2 G_0\tau_{d,eq}^2(\vec{\kappa}^\intercal\cdot\tens{D} + 
	\tens{D}\cdot\vec{\kappa}),
\end{align}
where we have neglected isotropic terms.  The stress is unaffected by disentanglement to $O(\dot\gamma^2)$; \new{the effects of CCR (\textit{i.e.} $\beta$)} will enter only at $O(\dot\gamma^3)$.

In steady-state shear with shear-rate $\vec{\kappa}=\dot\gamma\vec{e}_x\vec{e}_y$, we find
\begin{subequations}
	\begin{align}
		\sigma_{xy} &= G_0\tau_{d,eq}\dot\gamma + O({\dot\gamma}^3),\\
		\sigma_{xx} &= G_0(2-\alpha)\tau_{d,eq}^2{\dot\gamma}^2 + O({\dot\gamma}^4),\\
		\sigma_{yy} &= -G_0\alpha\tau_{d,eq}^2{\dot\gamma}^2 + O({\dot\gamma}^4), \\
		\nu &= 1 - \frac{\beta}{3}\tau_{d,eq}\tau_\nu{\dot\gamma}^2 + O({\dot\gamma}^4).
	\end{align}
\end{subequations}
The stress tensor is of second-order fluid form, with viscosity $\eta=G_0\tau_{d,eq}$, first normal stress coefficient ${\Psi_1=2G_0\tau_{d,eq}^2}$,  and second normal stress coefficient ${\Psi_2 = -\alpha G_0\tau_{d,eq}^2}$.  These results allow us to define $\alpha$ constitutively as
\begin{equation}
\lim_{\dot\gamma\to0}\frac{\Psi_2}{\Psi_1} = -\frac{\alpha}{2}.
\end{equation}
Thus, $\alpha$ can be determined by the normal stress ratio at low shear-rates.  For most melts, $-\Psi_2/\Psi_1\sim 0.1\textrm{--}.3$,\cite{mp-21} meaning $\alpha\sim0.2\textrm{--}0.6$.  Note that rod climbing will occur at low shear-rates only when $\alpha<0.5$.\cite{mp-21}

In uniaxial extension with rate  $\vec{\kappa}=\dot\epsilon(\vec{e}_z\vec{e}_z-\vec{e}_r\vec{e}_r)$, the non-zero stress components are
\begin{subequations}
	\begin{align}
		\sigma_{zz} &= G_0\left[2 + 4 \left(1 - \alpha\right)\tau_{d,eq}\dot\epsilon\right]\tau_{d,eq}\dot\epsilon + O({\dot\epsilon}^3),  \\
		\sigma_{rr}  & = G_0\left[-1 +  \left(1 - \alpha \right)\tau_{d,eq}\dot\epsilon\right]\tau_{d,eq}\dot\epsilon + O({\dot\epsilon}^3), \\
		\nu  &= 1 - \beta\tau_\nu\tau_{d,eq}{\dot\epsilon}^2 + O({\dot\epsilon}^3).
	\end{align}
\end{subequations}
The extensional viscosity is 
\begin{equation}
    \eta_E=3 G_0 (1 + (1 - \alpha)\tau_{d,eq}\dot\epsilon)\tau_{d,eq} + O({\dot\epsilon}^2).
\end{equation}  
\new{Thus, the extensional viscosity \new{increases} with increasing extension rate when $\dot\epsilon\ll\tau_{d,eq}^{-1}$, as was observed in simulations of Kremer-Grest melts with $Z_{e,eq}\sim7$ to $17$.\cite{oar-18}}
\end{widetext}

\subsection{Steady-state disentanglement}\label{subsec:nu-analytical}
The disentanglement equation \eqref{eq:governing-equation-nu} predicts
\begin{equation}\label{eq:nu-analytical}
	\nu = \exp\left[-W(\beta\tau_\nu\tens{S}:\bnabla\vec{v})\right]
\end{equation}
at steady state, where the Lambert $W$ function is defined by $W(x)e^{W(x)} \equiv x$.  This expression is independent of the details of the relaxation dynamics in $\mathbb{M}^{AA}$ and provides a pathway for testing predictions via simulations in which one can directly compute both the tube orientation tensor $\tens{S}$ and the normalized entanglement number $\nu$.  The model predicts that all measurements of the orientation tensor and entanglements can be fit with a single value of $\beta\tau_\nu$.  In simple shear, \eqref{eq:nu-analytical} can be written as
\begin{equation}\label{eq:nu-analytical-shear}
	\nu = \exp\left[-W\left(\frac{\beta\tau_\nu}{\tau_R}\textrm{Wi}_RS_{xy}\right)\right],
\end{equation} 
where $S_{xy}$ is the shear component of the orientation tensor and $ \textrm{Wi}_R=\dot\gamma\tau_R$ is the Rouse Weissenberg number. In extensional flow one obtains
\begin{equation}\label{eq:nu-analytical-extension}
	\nu = \exp\left[-W\left(\frac{\beta\tau_\nu}{\tau_R}\textrm{Wi}_R^E(S_{zz}-S_{rr})\right)\right],
\end{equation}
where $ \textrm{Wi}_R^E=\dot\epsilon\tau_R$ is the extensional Weissenberg number, and $S_{zz}$ and $S_{rr}$ are the extensional and compressional components of the orientation tension, respectively.
 
\color{black}
\section{Numerical Results}\label{sec:results}
\color{black}
Here, we investigate the full behavior of the model.  The microstructural predictions of our model are evaluated in Sec.~\ref{subsec:microstructure}. We study the steady state predictions for the number of entanglements  in steady-state shear in Sec.~\ref{subsec:steady-state}.  In Sec.~\ref{subsec:im-comparison}, we compare our model with the Ianniruberto-Marrucci model.  Transient behavior is examined in Sec.~\ref{subsec:results-transient}.  The disentanglement during startup shear and re-entanglement following cessation of steady shear is examined, and our determination of $\tau_\nu$ is appraised. \new{Finally, we discuss the non-linear rheological signatures of our model in Sec.~\ref{subsec:nonlinear-rheology}.}

\subsection{Microstructural predictions}\label{subsec:microstructure}
\begin{figure}
	\centering
	\includegraphics[width=0.45\textwidth]{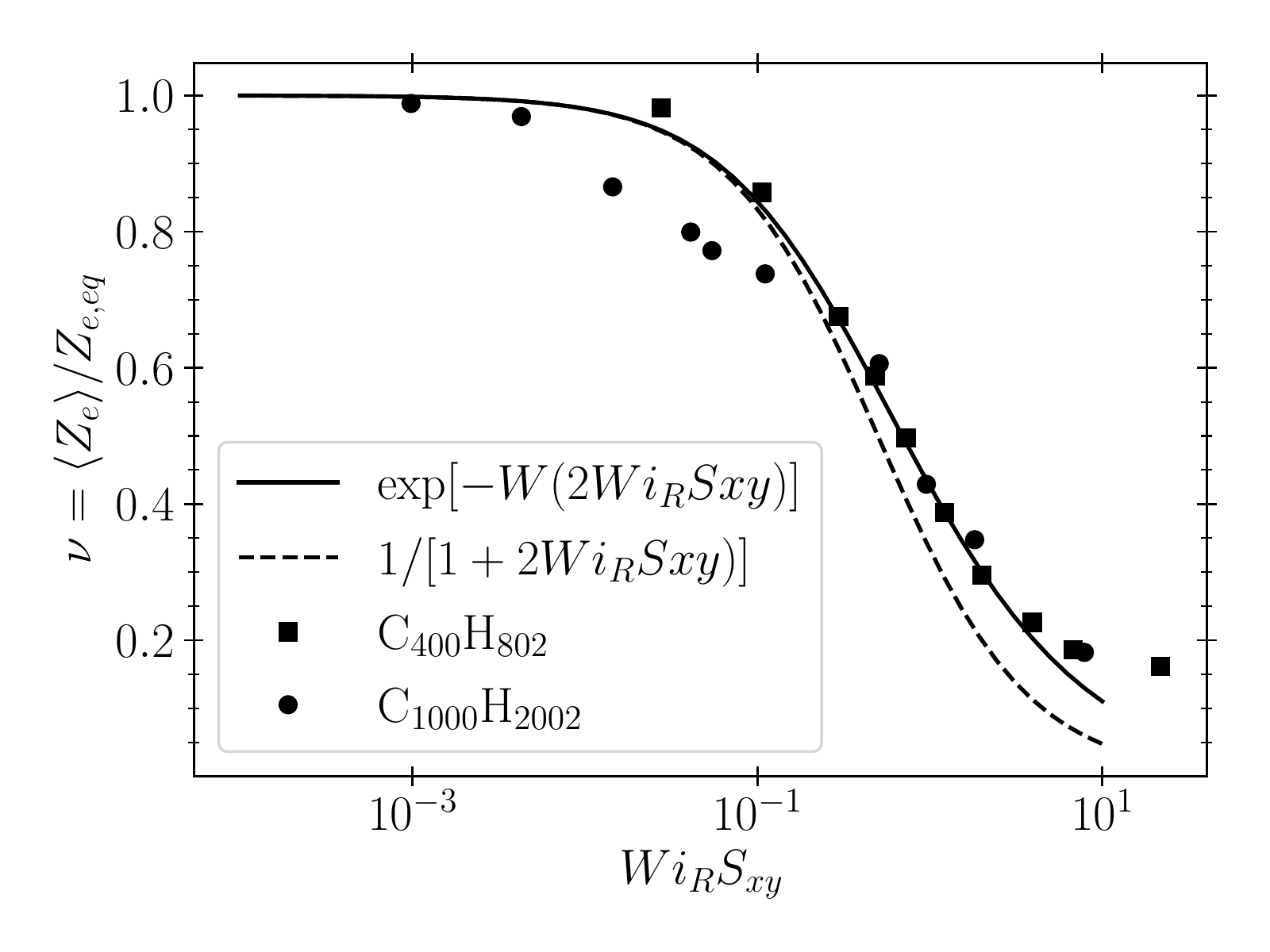}
	\caption{Comparison of the analytical prediction of $\nu$ given in \eqref{eq:nu-analytical} for steady-state shear flow to united atom simulations of $\textrm{C}_{400}\textrm{H}_{802}$\cite{bmk-10} and $\textrm{C}_{1000}\textrm{H}_{2002}$\cite{sek-19a,sek-19} polyethylene melts with \new{$\beta=2\tau_\nu/\tau_R$}.  The dashed line is the Pad\'e approximation of the analytical solution. }\label{fig:SxyTest}
\end{figure}

In figure~\ref{fig:SxyTest}, we compare the relation between the normalized entanglement number $\nu$ and orientation $\tens{S}$ (Eq~\ref{eq:nu-analytical-shear}) in steady shear flow to results from united atom simulations. \citet{bmk-10} simulated $\textrm{C}_{400}\textrm{H}_{802}$ polyethylene and computed $\nu$ and the conformation tensor $\tens{A}$; we compute the orientation tensor from their results using the relation $\tens{S}=\tens{A}/\tr\tens{A}$.  \citet{bmk-10} did not provide a value of $\tau_R$ so we use a value of $\tau_R$ computed in a different study of the same polymer.\cite{sek-15} \citet{sek-19a,sek-19} simulated $\textrm{C}_{1000}\textrm{H}_{2002}$ and computed the orientation tensor $\tens{S}$ directly.

The analytical expression \eqref{eq:nu-analytical-shear} fits both datasets with a single value of $\beta\tau_\nu/\tau_R=2$.  The value of $\beta$ that best fits the data thus depends on the ratio $\tau_\nu/\tau_R$.  In the I-M model, $\tau_\nu=\tau_{d,eq}$, which would imply $\beta\sim1/Z_{e,eq}$.  In our model, $\tau_\nu=\tau_R$, meaning $\beta$ is independent of molecular weight. This is consistent with the physical picture of the Rolie-Poly model, where $\beta$ is inversely proportional to the number of retraction events necessary to cause a tube hop on the order of the length of a tube segment.\cite{glmm-03,lg-03} Thus, we expect that $\beta$  depends on the length of a tube segment, and hence $N_{e,eq}$, but is independent of molecular weight.

\subsection{Shear-induced disentanglement}\label{subsec:steady-state}

We solve \eqref{eq:governing-equation-A} and \eqref{eq:governing-equation-nu} in steady-state simple shear in order to compute $\nu$.  Throughout this section we use the relationship $\tau_\nu=\tau_R$ found in Sec.~\ref{subsec:tau-nu}.  We begin by examining the influence of the maximum stretch $\lambda_\textrm{max}$ on disentanglement.  Using \eqref{eq:maximum-stretch-derived}, we find $\lambda_\textrm{max}\sim2$ for polyethylene, $\lambda_\textrm{max}\sim3\text{--}6$ for poly(ethylene–butene), where the lower(upper) bound corresponds to melts with few(many) ethyl branches, and $\lambda_\textrm{max}\sim9\text{--}11$ for poly(myrcene).\cite{PolymerHandbookChain}

\begin{figure}
	\centering
	\begin{minipage}[b]{.02\textwidth}
		(a) 
		\vspace{120pt}
	\end{minipage}
	\includegraphics[width=.43\textwidth]{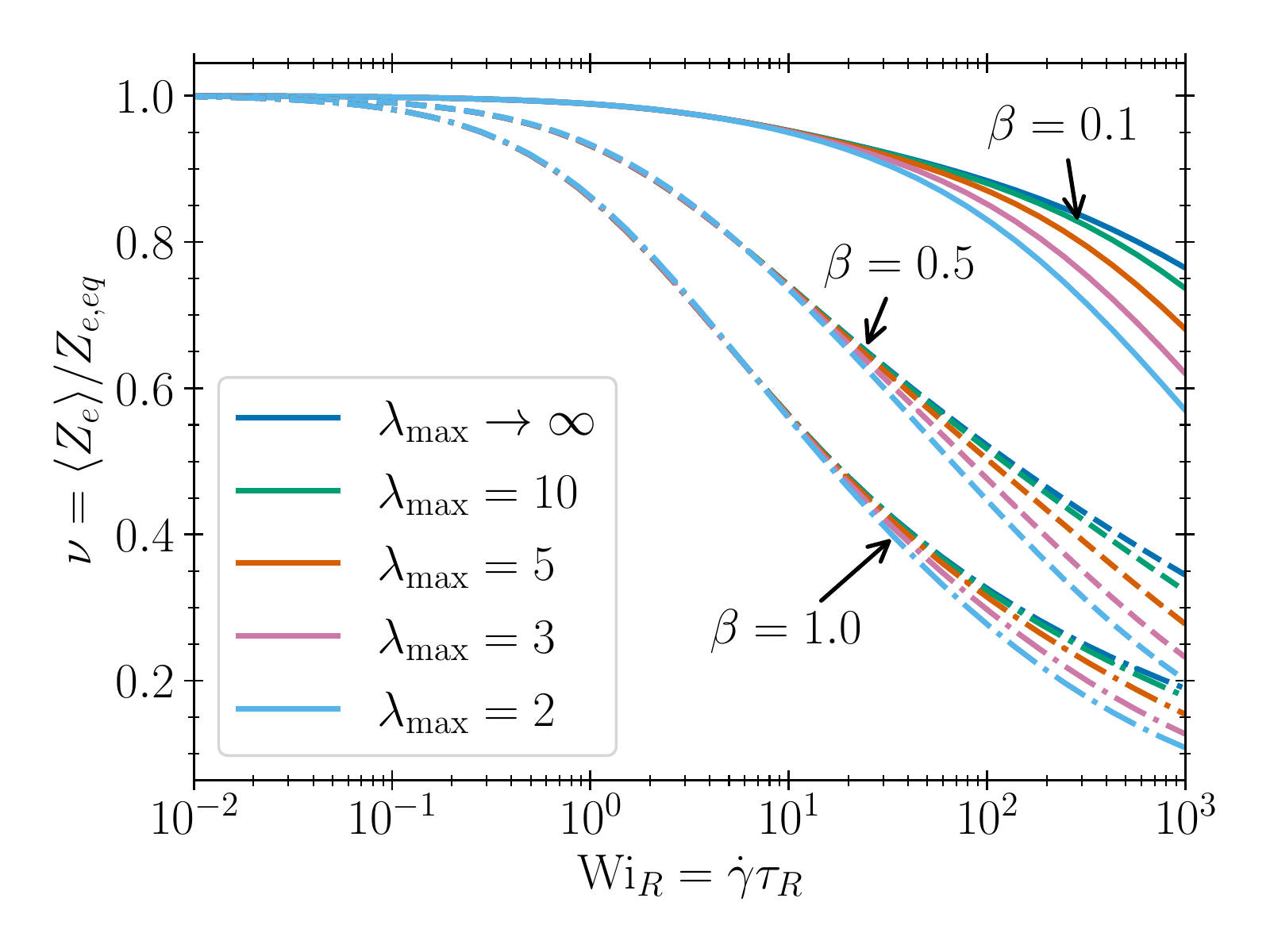}\\
	\begin{minipage}[b]{.02\textwidth}
		(b) 
		\vspace{120pt}
	\end{minipage}
	\includegraphics[width=0.43\textwidth]{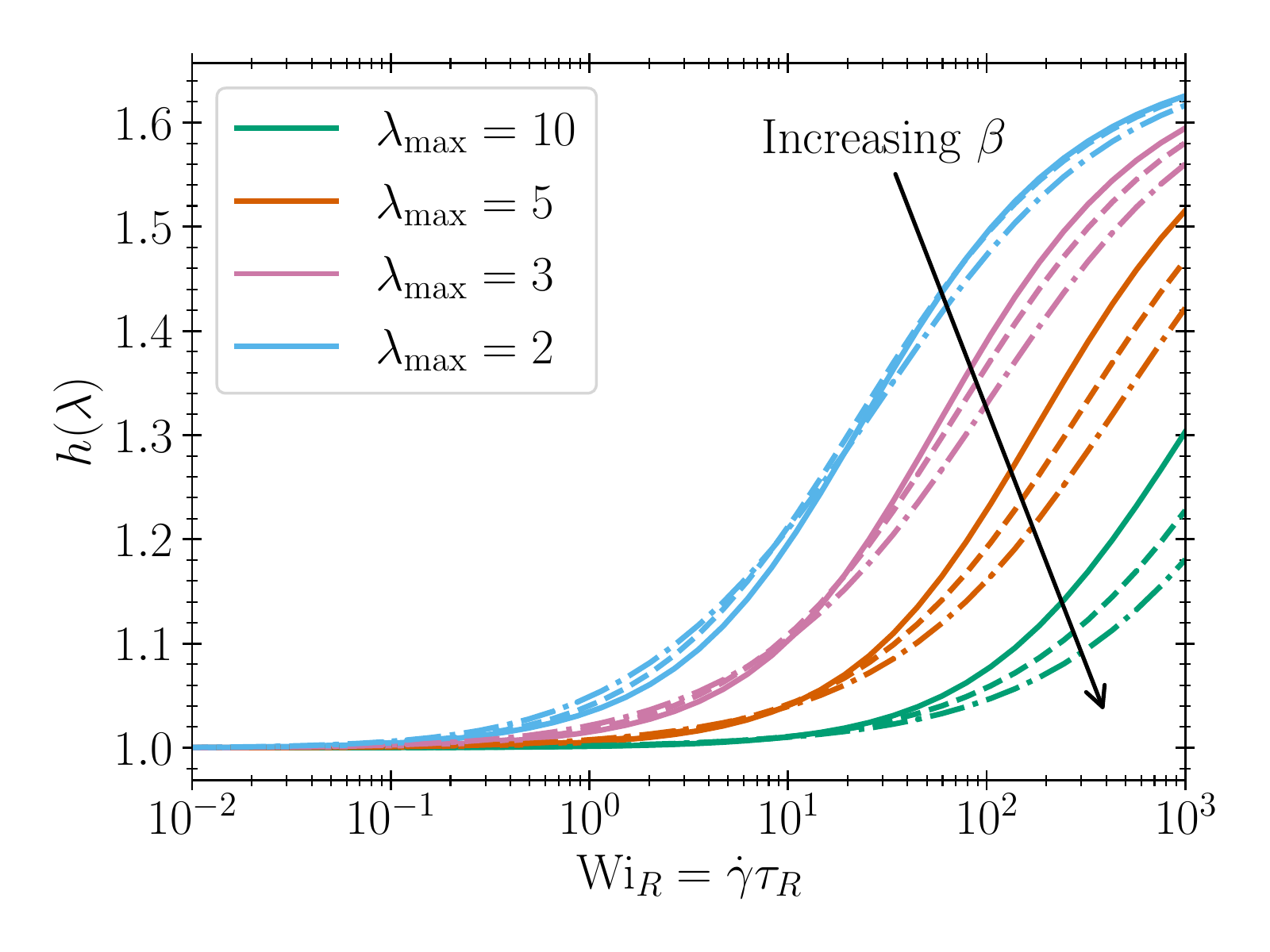}
	\caption{Influence of the maximum stretch on (a) the normalized entanglement number $\nu$ and (b) the spring force $\f(\lambda)$ in steady state shear flow.  Lines are obtained from solutions of \eqref{eq:governing-equations} with $Z_{e,eq}=50, \alpha=0.5$, $\beta=0.1,0.5, 1.0$ \new{(solid, dashed, dot-dashed)}, and $\tau_\nu=\tau_R$.}\label{fig:LambdaSweep}
\end{figure}

 The disentanglement of a melt with $Z_{e,eq}=50$ and $\alpha=0.5$ is plotted as a function of $\textrm{Wi}_R=\dot\gamma\tau_R$ for $\lambda_\textrm{max}=2,3,5,10$ and three values of $\beta=0.1,0.5,1.0$  in figure~\ref{fig:LambdaSweep}a.  The amount of disentanglement is independent of $\lambda_\textrm{max}$ for small $\textrm{Wi}_R$, as follows from the weakly non-linear analysis in \eqref{eq:nu-weakly-nonlinear}.  As the flow rate is increased the maximum stretch is approached, leading to a spring force $\f(\lambda)>1$ as shown in figure~\ref{fig:LambdaSweep}b.  The spring force is higher at a given $\textrm{Wi}_R$ for smaller $\lambda_\textrm{max}$ because less extensible chains  approaches the maximum stretch at lower flowrates.  The larger spring force increases the magnitude of the CCR term in the kinetic equation for $\nu$ (\eqref{eq:generic-nu}).  Thus, at a given $\textrm{Wi}_R$ melts with smaller $\lambda_\textrm{max}$ should disentangle more readily.  The influence of $\lambda_\textrm{max}$ is more pronounced for smaller $\beta$, where significant disentanglement only occurs as the maximum stretch is approached.  Careful consideration of $\lambda_\textrm{max}$ is thus required for accurately fitting values $\beta$ to data in melts at very high $\textrm{Wi}_R$.

\begin{figure}
	\centering
	\includegraphics[width=0.45\textwidth]{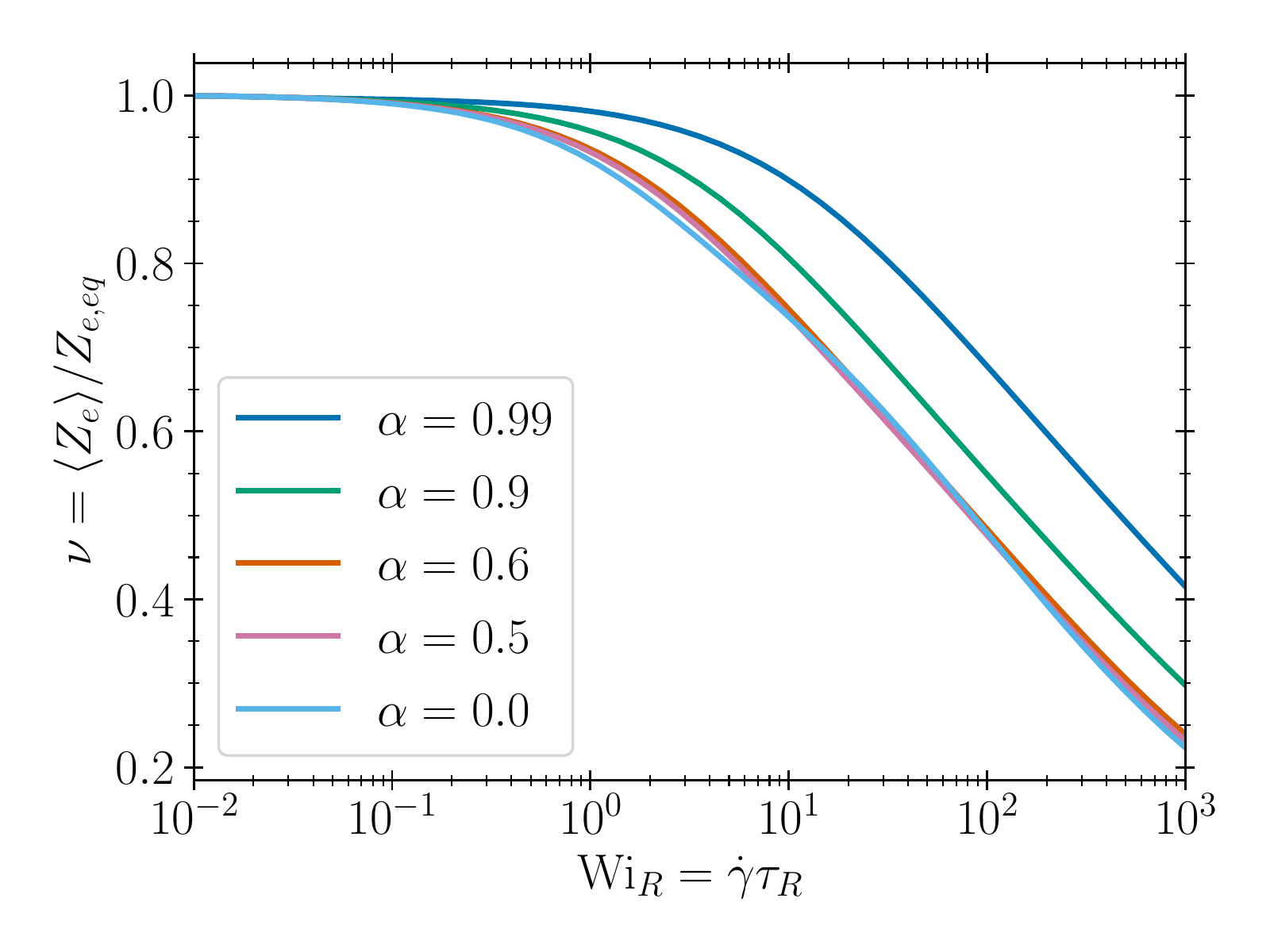}
	\caption{Influence of the \new{anisotropy parameter $\alpha$} on disentanglement in steady state shear according to  \eqref{eq:governing-equation-nu} with $Z_{e,eq}=50$, $\beta=0.5$, $\lambda_\textrm{max}=3$, and $\tau_\nu=\tau_R$}\label{fig:AlphaSweep}
\end{figure}

The \new{role} of the anisotropic mobility parameter $\alpha$ is examined in figure~\ref{fig:AlphaSweep} for $Z_{e,eq}=50$, $\beta=0.5$, and $\lambda_\textrm{max}=3$. 
For a larger  $\alpha$ the  chain mobility and relaxation rate increase, leading to a decrease in both the amounts of stretch and disentanglement.  However, this change is minimal until $\alpha\sim1$.  As discussed in Sec.~\ref{subsec:weakly-nonlinear}, the anisotropic mobility parameter $\alpha$ sets the value of the second normal-stress difference and has typical values $\alpha\sim0.2$ to $0.6$.  This implies that the second normal stress difference only weakly affects disentanglement for typical polymer melts.

\begin{figure}
	\centering
	\begin{minipage}[b]{.02\textwidth}
		(a) 
		\vspace{120pt}
	\end{minipage}
	\includegraphics[width=0.43\textwidth]{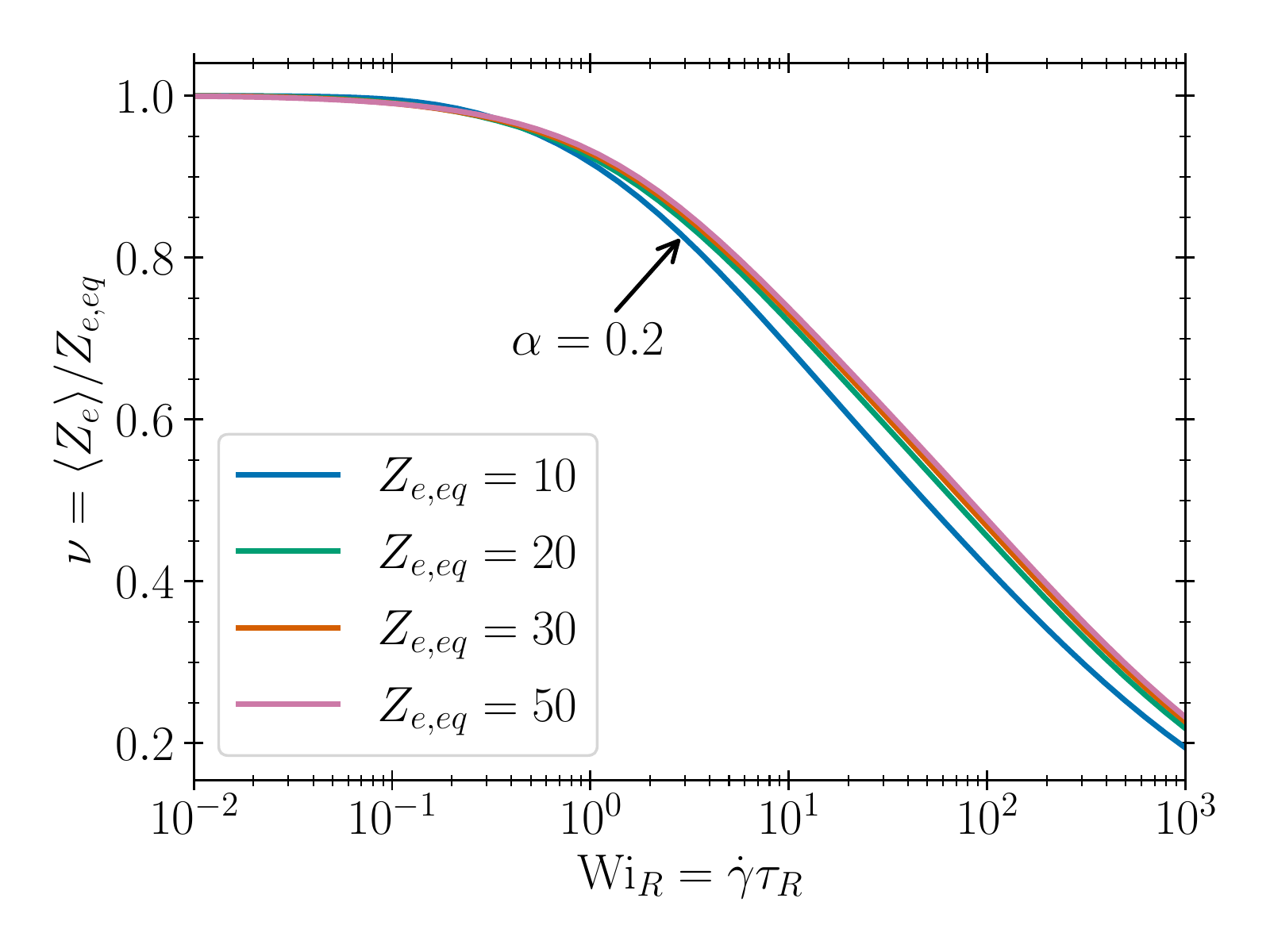}\\
	\begin{minipage}[b]{.02\textwidth}
		(b) 
		\vspace{120pt}
	\end{minipage}
	\includegraphics[width=0.43\textwidth]{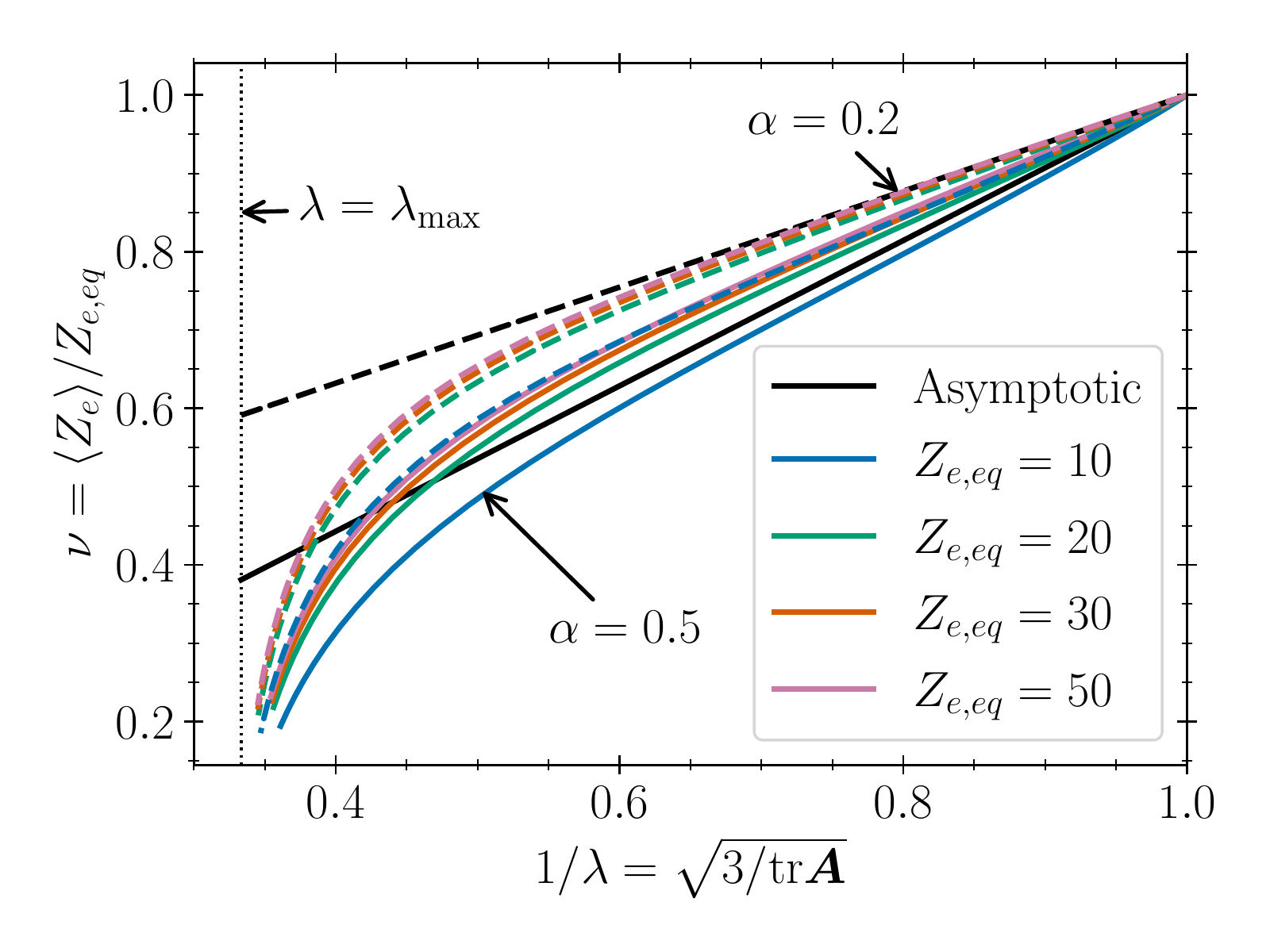}
	\caption{Influence of $Z_{e,eq}$ on disentanglement in steady state shear.  (a) Solutions of \eqref{eq:governing-equation-nu} for the normalized entanglement number $\nu$ with $\beta=0.5$, $\lambda_\textrm{max}=3$, $\alpha=0.5$, and  $\tau_\nu=\tau_R$. (b) Relationship between the stretch and entanglements in steady-state shear for $\alpha=0.2$ \new{(dashed curves)} and $\alpha=0.5$ \new{(solid curves)}. Small $\nu$ (left) corresponds to high $\Wi$ and $\lambda$, which increases towards equilibrium $(\Wi\rightarrow0,\lambda\rightarrow 1)$ on the right. The asymptotic results (solid and dashed lines) are obtained from \eqref{eq:nu-lambda-asymptotic}.}\label{fig:ZSweep}
\end{figure}

The influence of molecular weight \new{(\textit{i.e.} $Z_{e,eq}$)} is examined in figure~\ref{fig:ZSweep}a.  The normalized entanglement number in steady state shear is computed for $\beta=0.5$, $\lambda_\textrm{max}=3$, and $\alpha=0.5$, with $Z_{e,eq}=10,20,30$ and $50$.  The degree of disentanglement is nearly independent of molecular weight at fixed $\textrm{Wi}_R$,  because the re-entanglement rate and non-affine stretch rate, which are proportional to the inverse Rouse time, balance at steady state.  Thus, increasing the molecular weight increases $\tau_{d,eq}/\tau_R$ but has little influence on the stretch.  To confirm this we plot the normalized entanglement number as a function of $1/\lambda$ for $\alpha=0.2$ and $\alpha=0.5$ in figure~\ref{fig:ZSweep}b.  A linear relation is seen between $\nu$ and $1/\lambda$ for $0.6\alt 1/\lambda\alt 1$, or equivalently $1<\lambda<1.7\simeq\sqrt{\lambda_{\textrm{max}}}$.  The relationship between $\nu$ and $\lambda$ is largely independent of molecular weight for $Z_{e,eq}>10$.  Indeed, we find from \eqref{eq:A-weakly-nonlinear} and \eqref{eq:nu-weakly-nonlinear} that
\begin{subequations}
\begin{align}
    \nu&\simeq 
        1 - \frac{3\beta(\f^\prime(1)+2)}{6(1-\alpha) + 2\beta^2} \left(1-\frac{1}{\lambda}\right) + O\left(\frac{\tau_R}{\tau_d}\right)\label{eq:nu-lambda-asymptotic}\\
        &\simeq 1 - \frac{\beta(3N_{e,eq}-1)}{(N_{e,eq}-1)\left[3(1-\alpha)+ \beta^2\right]} \left(1-\frac{1}{\lambda}\right) 
\end{align}
\end{subequations}
near equilibrium $(\nu=1,\lambda=1)$.  This asymptotic result provides a reasonable approximation to the full numerical result for moderate stretching (dashed line in figure~\ref{fig:ZSweep}b).  A similar dependence of disentanglement on stretch and molecular weight was observed in molecular dynamics simulations,\cite{sek-19} suggesting that $\beta$ should be independent of molecular weight in our model.

\subsection{Comparison with the Ianniruberto-Marrucci model}\label{subsec:im-comparison}
\begin{figure}
	\centering
	\includegraphics[width=0.45\textwidth]{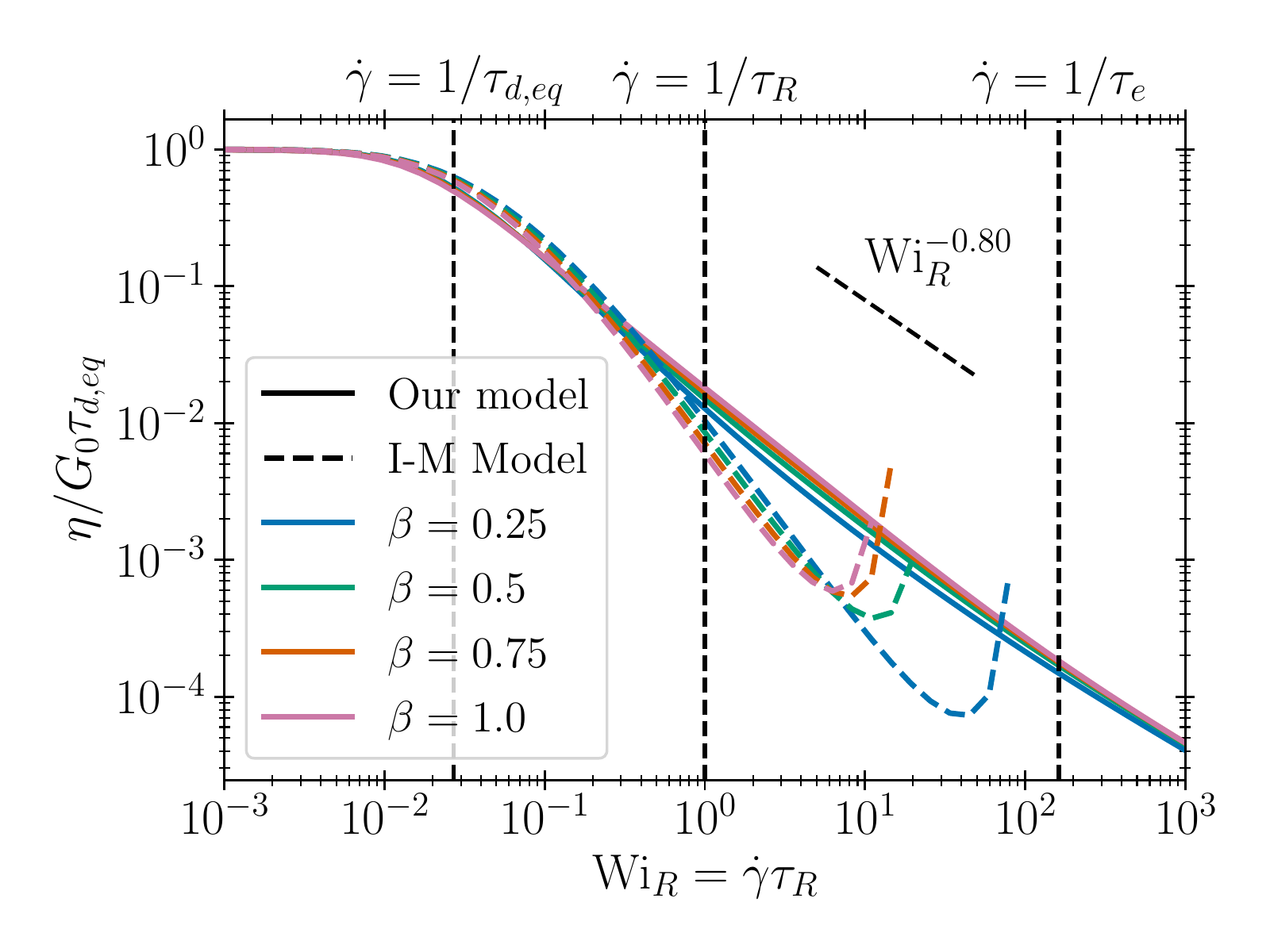}
	\caption{\new{Steady state viscosity $\eta$ with $Z_{e,eq}=50$, $\lambda_\textrm{max}=3$, $\alpha=0.5$, and $\tau_\nu=\tau_R$; plotted as a function of $\textrm{Wi}_R$.  Solid lines represent our model, dashed lines are the I-M model.}}\label{fig:BetaSweepEta}
\end{figure}

To compare predictions of our model to those of the I-M model we fixed the parameters $Z_{e,eq}=50$, $\lambda_\textrm{max}=3$, and $\alpha=0.5$.  Four values of $\beta=0.25,0.5,0.75$ and $1.0$ are examined.  Predictions of shear viscosity $\eta$ from our model (solid lines) and the I-M model (dashed lines) as a function of $\textrm{Wi}_R$ are plotted in figure~\ref{fig:BetaSweepEta}.  Both models approach the same value of $\eta$ in the small $\textrm{Wi}_R$ limit and shear-thin in stronger flows.  The I-M model predicts a lower viscosity at high shear rates, and contains a singularity at yet higher shear rates.  The origin of this behavior is found in the steady-state analytical solution of the kinetic equation for the stretch $\lambda_{\textrm{I-M}}$ \eqref{eq:im-stretch-relaxation} in the I-M model, 
\begin{equation}\label{eq:im-stretch-steady-state}
	\lambda_{\textrm{I-M}} = \frac{\nu^{1/2}}{1 - \textrm{Wi}_RS_{xy}},
\end{equation}
which is singular as $\textrm{Wi}_RS_{xy}\to1$.  This  singularity could be removed by using a FENE  multi-mode stretch equation proposed by \citet{ianniruberto-15} instead of \eqref{eq:im-stretch-relaxation}.  Note that the assumption that $\lambda$ equilibrates to $\nu^{1/2}$ leads to a decrease in stretch relative to that of a full-entangled melt.

Interestingly, the effects of $\beta$ on the viscosity are different between the two models.  In the I-M model, increasing $\beta$ leads to a faster effective reptation time at a given polymer stretch, reducing polymer alignment with the flow and hence reducing the viscosity.  This behavior is also present in the Rolie-Poly equation, which forms the basis of our model.  However, there is a second, competing effect in our model.  The cross coupling terms in \eqref{eq:governing-equation-A} increase the effective relaxation time of the conformation tensor because $\ln\nu<0$; increasing $\beta$ thus increases the magnitude of this effect.  The net effect is a small increase in viscosity with increasing $\beta$.

Prior work with the Rolie-Poly model\citep{lg-03,sek-19} and the Iannirubero-Marruci model\cite{im-14} suggested that $\beta\ll1$ in order to maintain good agreement with step strain experiments,\citep{lg-03} nonlinear rheology of united atom simulations,\citep{sek-19} and steady-state disentanglement in shear.\citep{im-14} In contrast, $\beta$ has a small influence on the rheological properties of our model, suggesting that it need not be small.  This is corroborated by tests of the microstructural predictions of our model presented in Sec.~\ref{subsec:microstructure}.

\begin{figure}
	\centering
	\begin{minipage}[b]{.02\textwidth}
		(a) 
		\vspace{110pt}
	\end{minipage}
	\includegraphics[width=0.43\textwidth]{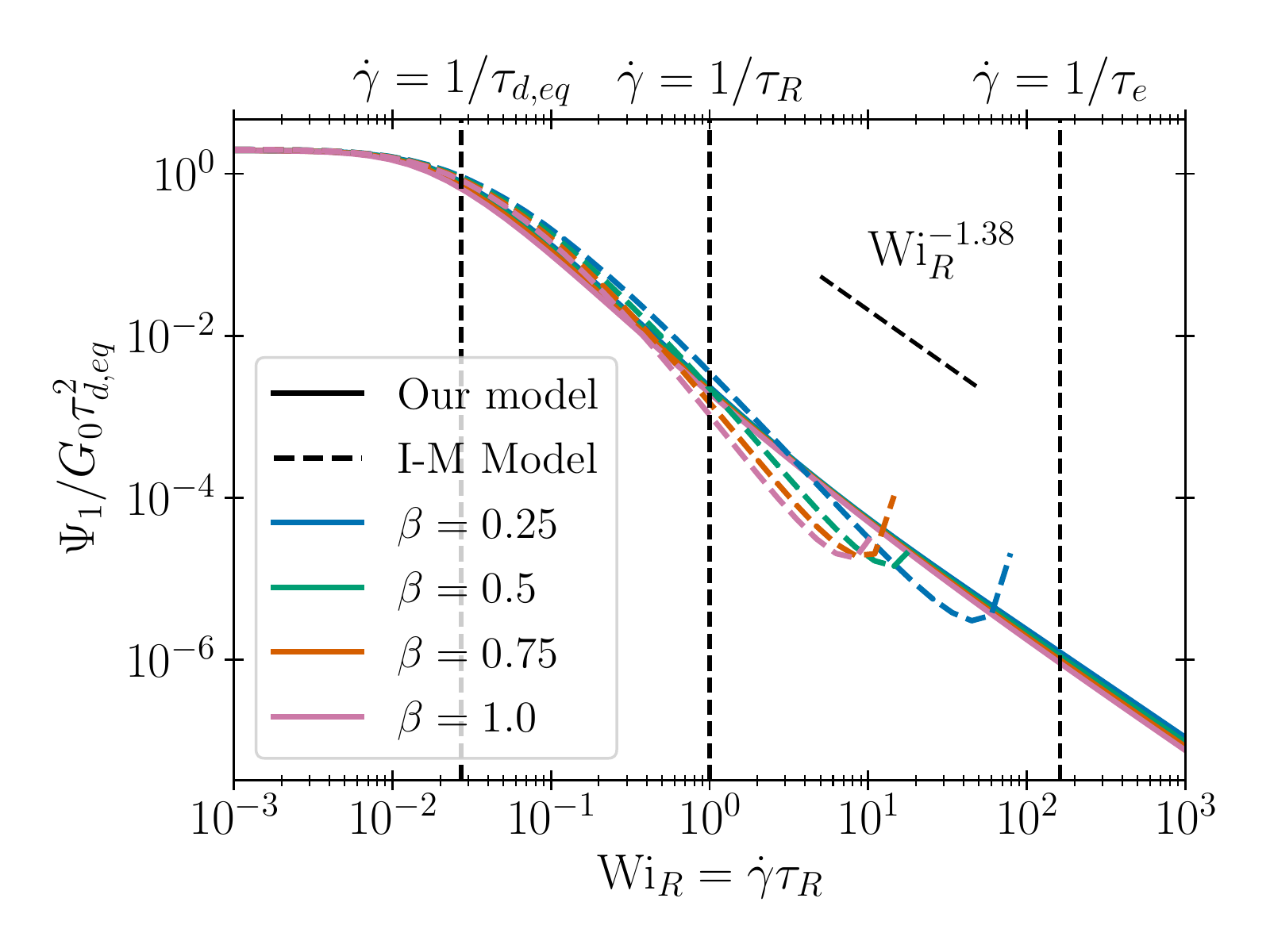}\\
	\begin{minipage}[b]{.02\textwidth}
		(b) 
		\vspace{110pt}
	\end{minipage}
	\includegraphics[width=0.43\textwidth]{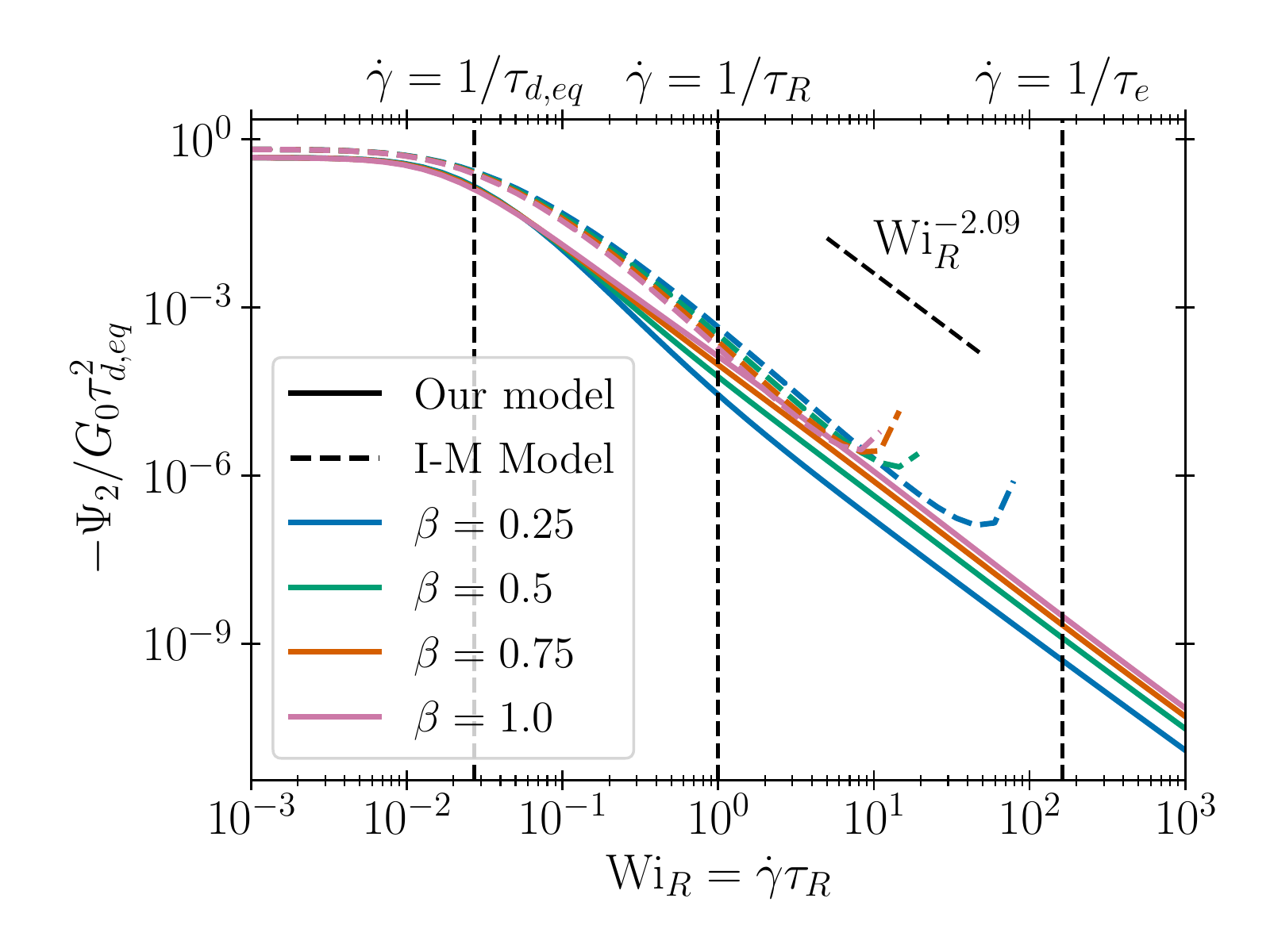}
	\caption{\new{Steady state first and second normal stress coefficients (a) $\Psi_1$ and (b) $\Psi_2$ with $Z_{e,eq}=50$, $\lambda_\textrm{max}=3$, $\alpha=0.5$, and $\tau_\nu=\tau_R$; plotted as a function of $\textrm{Wi}_R$.  Solid lines represent our model, dashed lines are the I-M model.}}\label{fig:BetaSweepPsi}
\end{figure}

Predictions of the first and second normal-stress coefficients $\Psi_1$ and $\Psi_2$ as a function of $\textrm{Wi}_R$ are shown in figure~\ref{fig:BetaSweepPsi}.  In each model, both normal-stress coefficients decrease with increasing shear rates.  In the slow flow limit both models predict the same values of $\Psi_1$.  For the I-M model (dashed lines) the trends observed in $\Psi_1$ are identical to those observed in $\eta$.  The parameter $\beta$ in our model (solid lines) has a weaker influence on $\Psi_1$ than on $\eta$.  Furthermore, increasing $\beta$ decreases the first normal stress coefficient.  This behavior can be attributed to \new{the} anisotropic mobility \eqref{eq:Giesekus}, which increases the effective reptation rate in the direction of polymer alignment with the flow.  This counteracts the cross coupling terms in \eqref{eq:governing-equation-A}, leading to a slight decrease in $\Psi_1$ with increasing $\beta$.  

Both models predicts qualitatively similar behaviors for $\Psi_2$ at low shear rates, albeit with slightly different plateau values of $\Psi_2$.  In the I-M model the definition of the deformation measure \eqref{eq:Q} fixes the low shear rate normal stress ratio to $\Psi_1/\Psi_2=-1/3$, which is somewhat larger in magnitude than the values of $-\Psi_1/\Psi_2\simeq0.1\textrm{-} 0.3$ typically measured in experiments.\cite{mp-21}  In  our model the ratio  $\Psi_1/\Psi_2=-\alpha/2$ is set by the anisotropic mobility parameter $\alpha$, and so can be treated as a measurable constitutive parameter.  Both models predict a negative second normal stress coefficient that monotonically decreases with increasing flow rate.  Increasing $\beta$ increases the magnitude of $\Psi_2$ in our model.  The influence of $\beta$ on $\Psi_2$ is more noticeable than on $\eta$ and $\Psi_1$.  This behavior is largely because the second normal-stress coefficient is much smaller than the viscosity and the first normal-stress coefficient.

\new{The viscosity and first and second normal stress coefficients (\textit{c.f.} figures \ref{fig:BetaSweepEta} and \ref{fig:BetaSweepPsi}) exhibit power law behavior independent of $\beta$ at high $\Wi_R$.  We obtain exponents of $-0.80$,$-1.38$, and $-2.09$, for $\eta$, $\Psi_1$, and $\Psi_2$, respectively by fitting over the range $\Wi_R=10$ to $1000$.  The power-law exponent for viscosity agrees well with the value $\eta\sim\Wi_R^{-0.82}$ measured in united-atom simulations of polyethylene simulations over the range $\Wi_R\geq3$.  Those same simulations found that {$\Psi_1\sim\Wi_R^{-1.75}$}, and $\Psi_2\sim\Wi_R^{-1.95}$.  Both $\Psi_1$ and $\Psi_2$ strongly thin in our model and in the simulations, although the values of the thinning exponents differ somewhat.}

\begin{figure}
	\centering
	\includegraphics[width=0.45\textwidth]{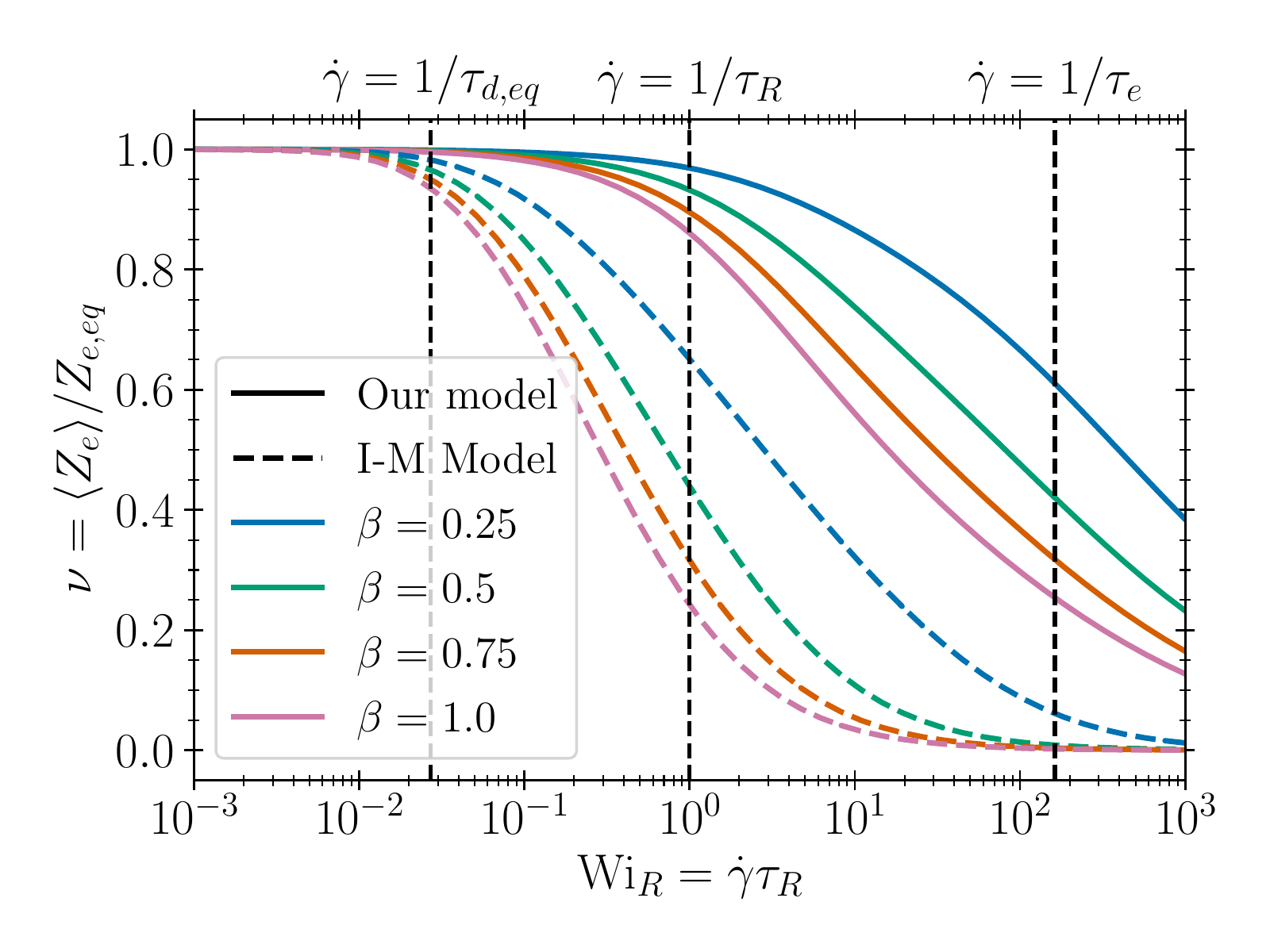}
	\caption{The effect of $\beta$ on the normalized entanglement number $\nu$  as a function of $\textrm{Wi}_R$ in  steady state shear flow, for $Z_{e,eq}=50$, $\lambda_\textrm{max}=3$, $\alpha=0.5$, and $\tau_\nu=\tau_R$.  Solid lines are predictions of our model and dashed lines are the I-M model.}\label{fig:BetaSweepNu}
\end{figure}

The normalized entanglement number $\nu$ obtained from \new{the} solution of \eqref{eq:governing-equations} is plotted in figure~\ref{fig:BetaSweepNu}.  The overall effect of $\beta$ is to increase the amount of disentanglement.  The I-M model (dashed lines) predicts a \new{larger degree of disentanglement at a given value of $\Wi_R$}.  This is largely due to the slower rate of re-entanglement, governed by $\tau_\nu=\tau_{d,eq}$, in the  I-M model\cite{im-14}.  Our model assumes that the melt re-entangles on $\tau_\nu=\tau_R$, which leads to less of disentanglement for a given $\beta$.  Hence, the value of $\beta$ that best fits a given experiment actually depends on the details of the model, in contrast with $Z_{e,eq}$, $\tau_{d,eq}$, $\tau_R$, $\lambda_\textrm{max}$, and $\alpha$, which can all be measured or computed independently.

\subsection{Kinetics of (dis)entanglement}\label{subsec:results-transient}
\begin{figure}
	\centering
	\begin{minipage}[b]{.02\textwidth}
		(a) 
		\vspace{110pt}
	\end{minipage}
	\includegraphics[width=0.43\textwidth]{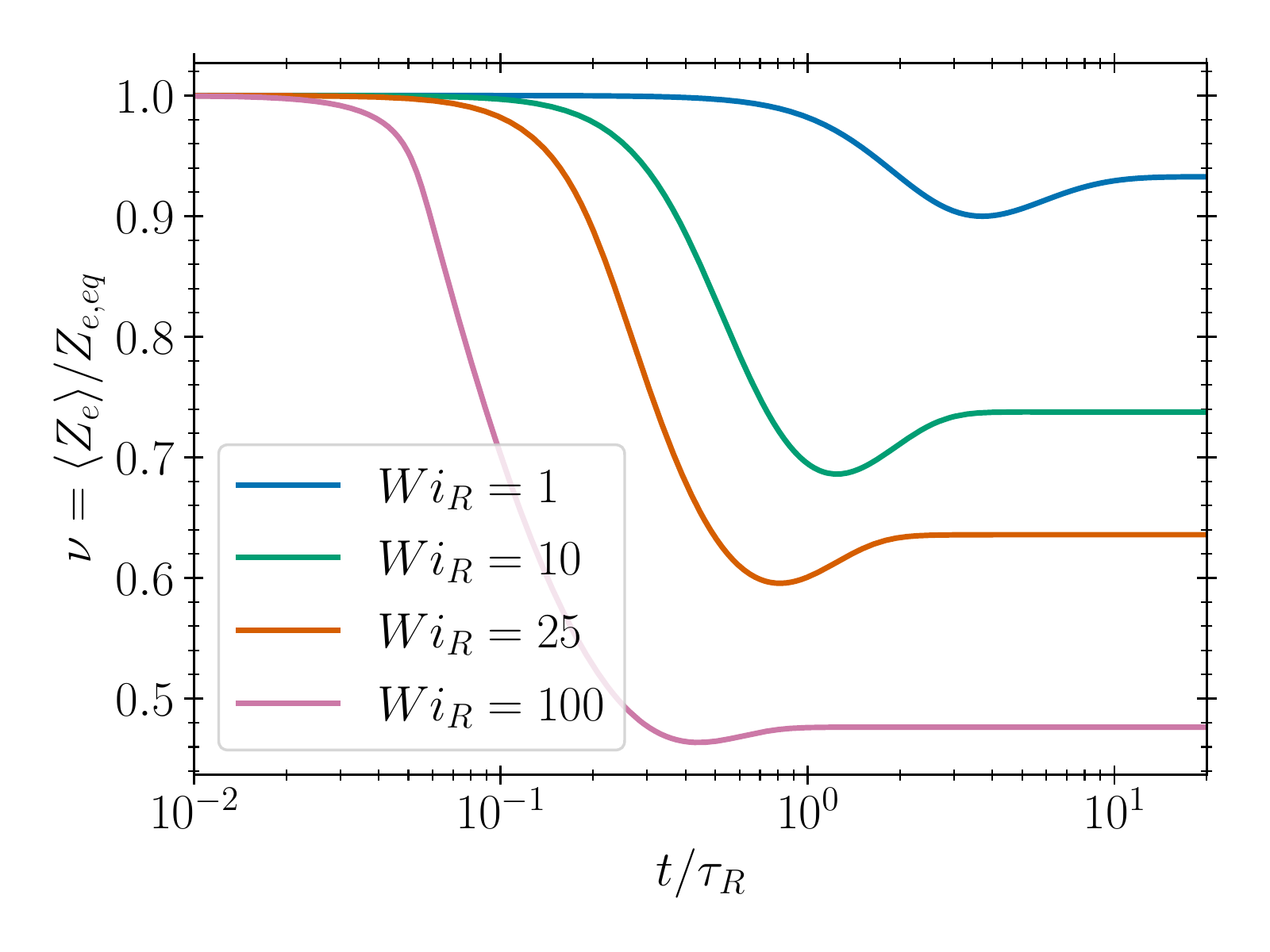}\\
	\begin{minipage}[b]{.02\textwidth}
		(b) 
		\vspace{110pt}
	\end{minipage}
	\includegraphics[width=0.43\textwidth]{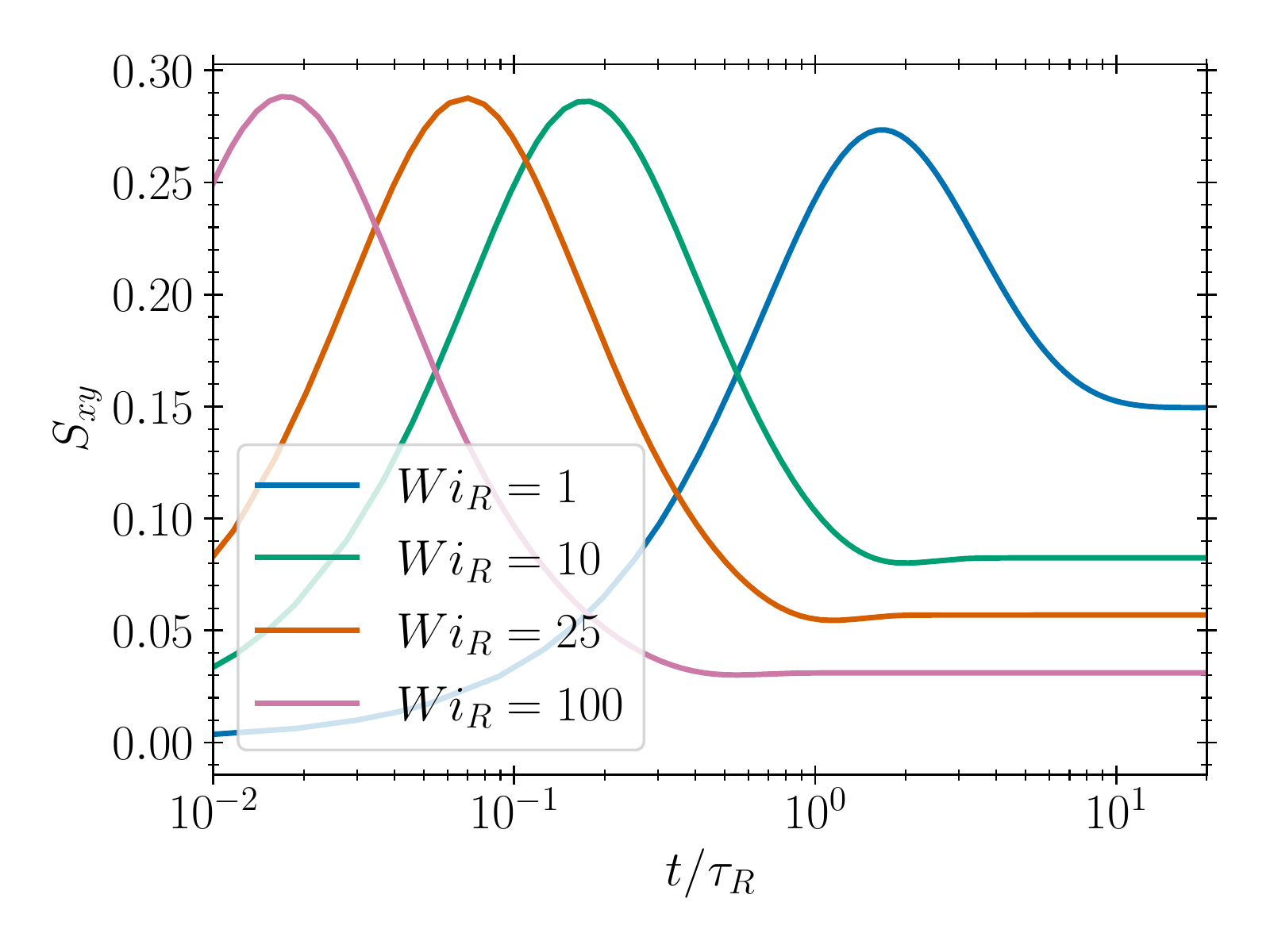}
	\caption{Evolution of (a) $\nu$ and (b) $S_{xy}$ during startup shear for a melt with $Z_{e,eq}=50$, $\beta=0.5$, $\lambda_\textrm{max}=3$, $\alpha=0.5$, and $\tau_\nu=\tau_R$.}\label{fig:StartupNu}
\end{figure}


Transient disentanglement during startup shear for a melt with $Z_{e,eq}=50$, $\beta=0.5$, $\lambda_\textrm{max}=3$ and $\alpha=0.5$ is plotted in figure~\ref{fig:StartupNu}a as a function of time.  \new{Upon startup, polymer chains will slip past each other, removing entanglements from the chain ends.  At longer times, the tube orientation in the shear directions decreases (figure~\ref{fig:StartupNu}b), decreasing the rate of disentanglement and leading to an undershoot of the steady-state number of entanglements.  This physical picture is mostly consistent with molecular dynamics simulations.  In simulations, the evolution of entanglements in startup shear flow exhibits an overshoot above the steady-state number of entanglements followed by an undershoot of the steady-state value.\cite{sek-19a,gor-22}  The undershoot in entanglements in our model lies between the undershoot and overshoot times of the shear orientation (figure~\ref{fig:StartupNu}b) in agreement with molecular dynamics simulations.\cite{sek-19a}   Our model correctly predicts the undershoot, but does not describe the overshoot.  It has been suggested that a correct description of the overshoot requires constitutive equations that account for directional correlations between entanglements.\cite{gor-22}} 

\begin{figure}
	\centering
	\includegraphics[width=0.45\textwidth]{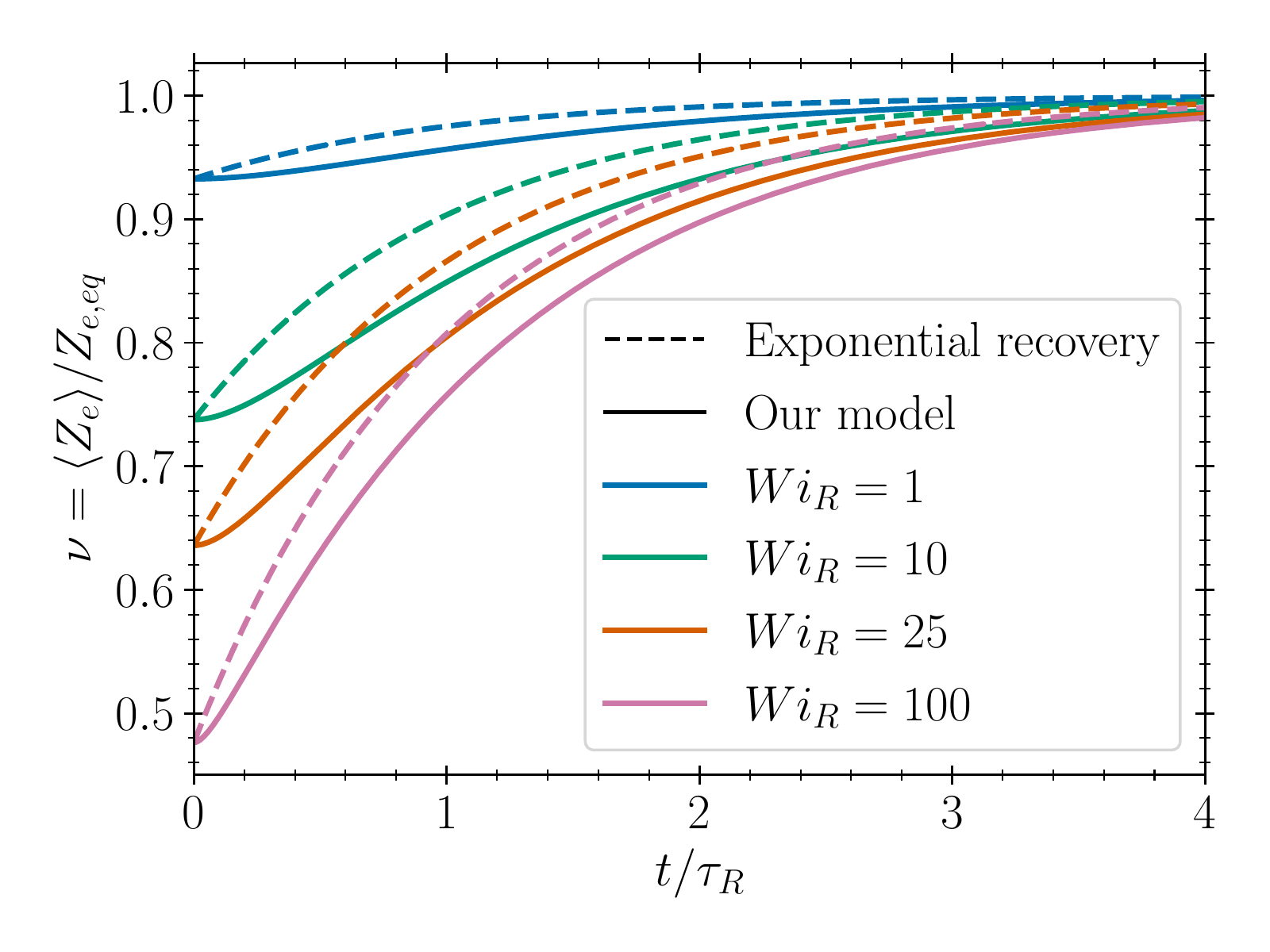}
	\caption{Re-entanglement following cessation of a steady state shear flow with  $Z_{e,eq}=50$, $\beta=0.5$, $\lambda_\textrm{max}=3$, $\alpha=0.5$, and $\tau_\nu=\tau_R$.  Solid lines are our model and the dashed line represents exponential recovery on \new{a timescale $\tau_\nu=\tau_R$.}}\label{fig:CessationNu}
\end{figure}
Re-entanglement of a melt following cessation of steady state shear is plotted in figure~\ref{fig:CessationNu}.  The re-entanglement of the melt is monotonic in time and approximately exponential in $\tau_\nu$, but is slightly slowed by the removal of entanglements at the chain ends as the stretch relaxes.  Indeed, the re-entanglement rate would be faster than exponential in the absence of this mechanisms due to the logarithmic form of the re-entanglement term in  \eqref{eq:governing-equation-nu}.  The removal of entanglements is captured mathematically by the negative non-affine stretch rate in \eqref{eq:governing-equation-nu} during chain retraction.

\new{
\subsection{Non-linear rheology}\label{subsec:nonlinear-rheology}
\begin{figure}
    \centering
    \includegraphics[width=0.45\textwidth]{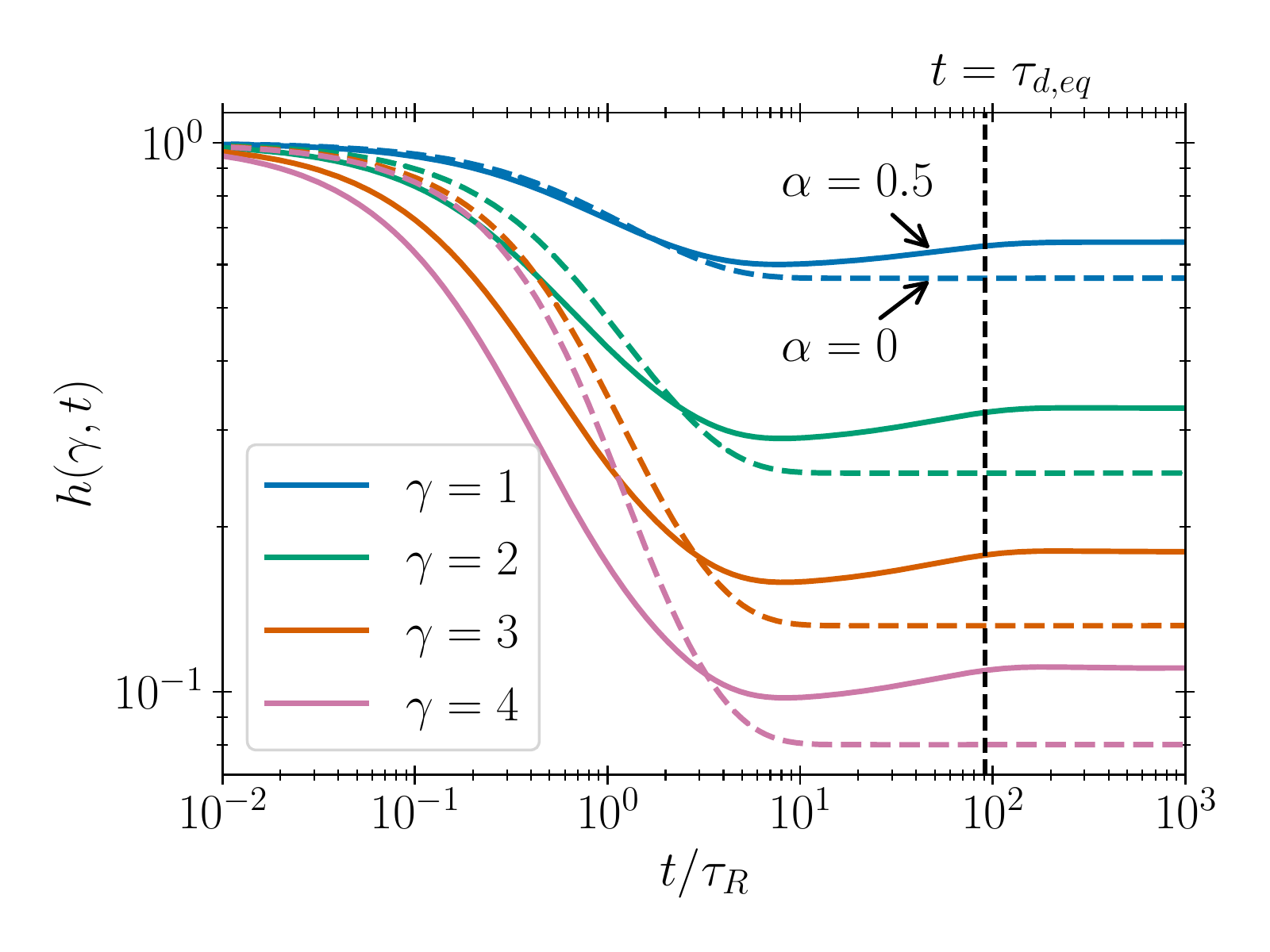}
    \caption{\new{Influence of the anisotropic mobility parameter $\alpha$ on the transient damping function $h(\gamma,t).$  Lines are obtained from solutions of \eqref{eq:governing-equations} for relaxation following an imposed strain of magnitude $\gamma$ with $Z_{e,eq}=100$, $\beta=1$, $\lambda_\textrm{max}\to\infty$, $\alpha=0,0.5$ (solid, dashed), and $\tau_\nu=\tau_R$.}}
    \label{fig:DampingFunction}
\end{figure}
We investigate the non-linear rheological behavior of our model, and show that experimental signatures previously interpreted as indicative of re-entanglement on the reptation time are consistent with $\tau_\nu=\tau_R$.  We solve our model for the transient relaxation following a step strain of magnitude $\gamma$, and plot the transient damping function
\begin{equation}
    h(\gamma,t) = \frac{G(\gamma,t)}{G(t)}
\end{equation}
in \ref{fig:DampingFunction}, where $G(\gamma,t)$ is the non-linear relaxation modulus and the linear relaxation modulus is given by \eqref{eq:linear-modulus}.  Time-strain separability occurs for all values of the anisotropic mobility $\alpha$ parameter considered here, as indicated by the presence of a long-time plateau in the damping function.  However, the timescale where time-strain separability is reached is dependent on $\alpha$.  For $\alpha=0$, the damping function decays on the Rouse time before plateauing at a time comparable to $\tau_R$.  In contrast, the damping function exhibits an undershoot when $\alpha=0.5$, and time-strain separability is only reached after $\tau_{d,eq}$.  This later behavior is consistent with step-strain experiments in entangled polymer solutions.\cite{sa-02}

The I-M model also predicts that the transient damping function contains an undershoot and the that time-strain separability is reached on the reptation time.  In the I-M model these behaviors arise due to re-entanglement on the reptation time and the assumption that stretch relaxes to a length set by the current number of entanglements.  In our model, the observed dynamics arise from the non-linear relaxation arising from the anisotropic mobility.  Thus, time-strain separability on the reptation time is not a reliable experimental signature of slow re-entanglement on the reptation time.

\begin{figure}
    \centering
    \includegraphics[width=0.45\textwidth]{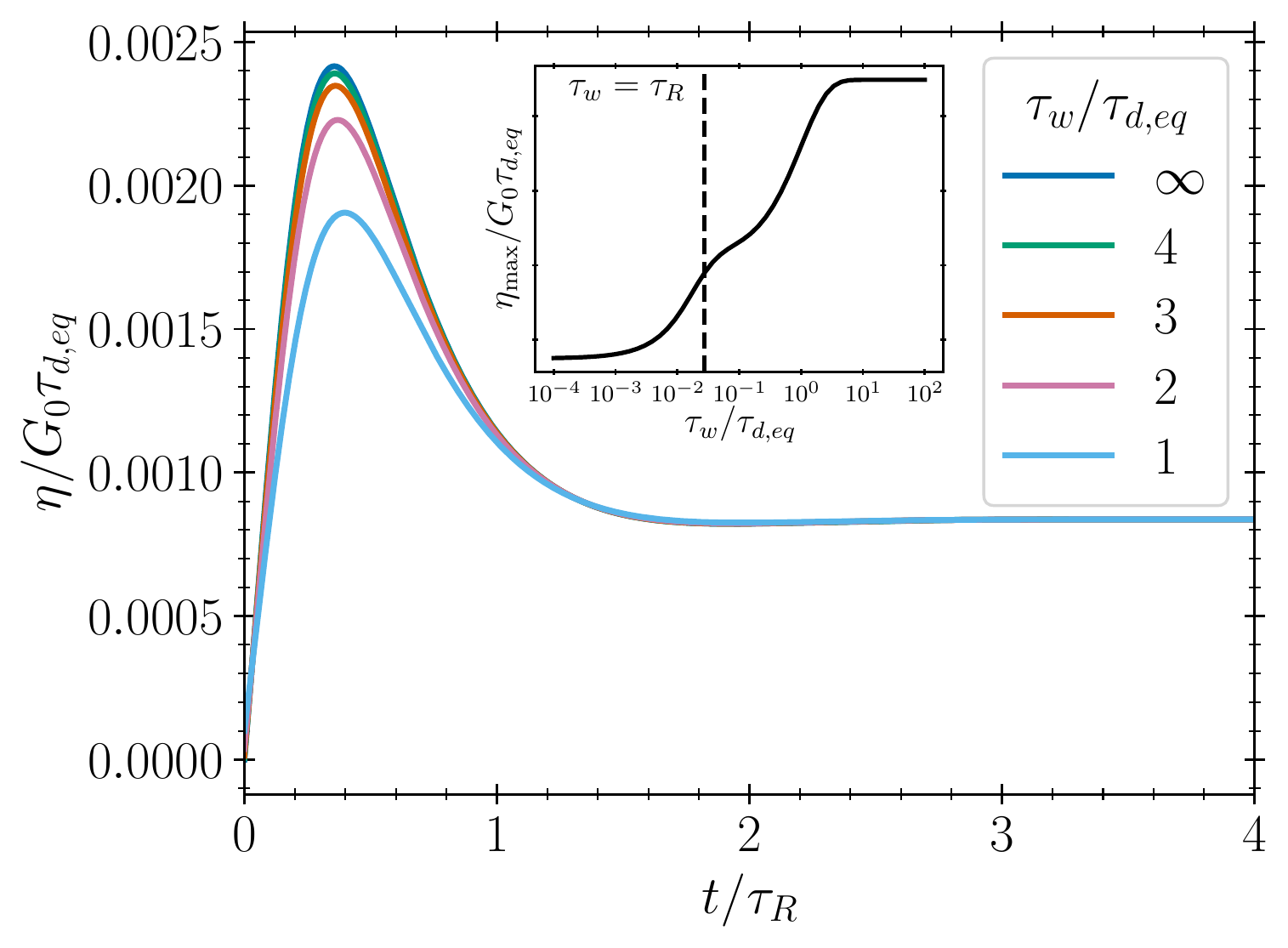}
    \caption{\new{Dependence of the viscosity overshoot on  waiting time $\tau_w$ at $\Wi_R=10.$  Lines are obtained from solutions of \eqref{eq:governing-equations} for $Z_{e,eq}=100$, $\beta=1$, $\lambda_\textrm{max}\to\infty$, $\alpha=0.5$, and $\tau_\nu=\tau_R$. Inset: Height of the overshoot as a function of waiting time.}}
    \label{fig:RepeatedShear}
\end{figure}

We plot our predictions for the transient viscosity during repeated startup shears in figure~\ref{fig:RepeatedShear}.  The melt is sheared to steady state at $\Wi_R=10.$  The flow is stopped and the melt relaxes for a waiting time $\tau_w$, after which the melt is sheared to steady state again at the same rate.  The first shear is given by $\tau_w\to\infty$.  The height of the overshoot during the second shear increases with increasing waiting time, eventually approaching the overshoot of the first shear for $\tau_w\sim4\tau_{d,eq}$.  In experiments, the waiting time required for the overshoot recovery has been interpreted as the timescale of re-entanglement.\cite{sb-73,rr-13}  Our results suggest that this interpretation is not correct; the overshoot recovery timescale in our model ($\tau_{d,eq}$) is much longer than the re-entanglement time ($\tau_\nu=\tau_R$).  It should be mentioned that contrary to our model and the tube model, the timescale of the overshoot recovery in experiments is often as many as orders of magnitude higher than the orientational relaxation time. \citet{im-14c} suggested that this discrepancy is due to polydispersity, and that the recovery of the overshoot is dominated by the longest (slowest) chains.  In support of this, we note that the entanglement recovery time in experiments is of the same order of magnitude as the longest relaxation time.\cite{rr-13}
}

\section{Conclusions}\label{sec:conclusion}
In this work, we have derived a thermodynamically consistent framework for describing the disentanglement of a polymer melt under flow. We provide three versions of this framework, with increasing degree of specificity:
\begin{enumerate}
    \item  Eqs.~\ref{eq:generic} are fully general, for any choice of the mobility tensors $\tens{M}^{\nu A}, M^{\nu\nu}$ and $\mathbb{M}^{AA}$, and free energy $F$. 
    \item Eqs.~(\ref{eq:generic-Mob}, \ref{eq:generaic-im}) implement a physical assumption that mobilities involving entanglement dynamics via convective constraint release arise from the same physical effects \new{(segmental motion)} as the  mobility that governs the stretch contained within the conformation tensor. This allows for a description entirely in terms of the mobility tensor $\mathbb{M}^{AA}$ \new{that} governs the relaxation of the conformation tensor. 
    \item Eqs.~\ref{eq:governing-equations} constitute a specific form for the dynamics, obtained by assuming a physically reasonable form for $\mathbb{M}^{AA}$ that separates into additive effects governing reptation and retraction. This choice reproduces the Ianniruberto-Marrucci mechanism for CCR, together with the dynamics of the conformation tensor implicit in the Rolie-Poly model, and incorporates finite elastic effects (FENE). Other dynamics may be obtained by suitable assumptions for the mobility tensors and the free energy.
\end{enumerate}

We have motivated these forms as arising from a Green-Kubo relation in which particular forms have been chosen for how the fast degrees of freedom (Appendix~\ref{app:green-kubo}) couple to dissipative retraction and reptation dynamics effects of the conformational tensor.  These noise choices ultimately arise from the same degrees of freedom (inter- and intra-chain dynamics faster than $\tau_e$).  Our choice is a plausibility argument whose justification is to obtain the phenomenological form of the I-M mechanism of CCR. We expect that this simple form is strictly incorrect, and it (and the full equation set) will fail for more complex and  time-dependent flows.

Hence, the disentanglement rate is governed by the I-M\cite{im-14} disentanglement mechanism, which assumes that the melt disentangles due to non-affine stretch of the molecules.   Rolie-Poly dynamics were used to specify the relaxation of the conformation tensor. We postulated a form for how the free energy changes with the number of entanglements.  Thermodynamic consistency requirements then lead to couplings between the kinetic equation for the melt conformation tensor,  the normalized entanglement number, and the total stress.  The overall effect of the new couplings is that increasing the convective-constraint release rate slightly increases the stress, and changes the form of the disentanglement in strong flows.

Prior theoretical work assumed that the melt re-entangles on the reptation time,\cite{im-14,im-14b,ianniruberto-15,hhhr-15,mbp-15,mmp-18} in disagreement with molecular dynamics simulations that predicted faster re-entanglement.\cite{ohr-19,Marco_Thesis,bsek-22}  We capture this fast re-entanglement by introducing a re-entanglement time $\tau_\nu=\tau_R$, which leads to melt re-entanglement before stress relaxation.  That is, entanglements `recover' by chain motion on the Rouse time, which brings the melt to a fully entangled and anisotropic state with residual stresses. The remaining stress then relaxes as the melt returns to the isotropic stress-free average configuration over the reptation time. \new{The equilibration dynamics of $\nu$ \textit{within} the tube is not accounted for in this (or any other) model, and may play an important role in understanding how the full equilibrium melt state recovers.}

In steady state, our model predicts an exact relationship between the orientation tensor and the number of entanglements.  This analytical solution agrees well with united atom simulations of polyethylene\cite{bmk-10,sek-19a,sek-19} at two different molecular weights.  These results, together with the assumptions $\tau_\nu=\tau_R$, suggest that $\beta$, the parameter that controls the rate of shear-induced disentanglement, is independent of molecular weight.  We find that finite extensibility plays an important role in melts with small $\beta$, and is less important in melts with large $\beta$.  Our model predicts a shear-thinning viscosity, a positive first normal stress difference, and a negative second normal stress difference.  We find that for the second normal stress difference does not influence shear-induced disentanglement \new{for physically relevant values of the anisotropic mobility paramter $\alpha$}.  Our model predicts that a melt will undershoot the steady-state number of entanglements during startup shear flow.  This undershoot is due to the stretch overshoot as the melt orients with the flow.  We find that the re-entanglement of a melt following cessation of steady-state shear is slightly slower than exponential recovery on the Rouse time. 

\new{We have shown that our model is consistent with transient nonlinear rheology experiments, demonstrating that both step-strain separability and recovery of the viscosity overshoot occur on the reptation time.  Prior work has interpreted these results as experimental signatures of re-entanglement on the reptation time.  Our results provide an alternative  interpretation, and suggest that the results may instead be interpreted as arising from anisotropic orientational relaxation.}

The convective-constraint release mechanism, as we have formulated it, arises from the assumption that the rate of disentanglement is proportional to the non-affine stretch rate.  Future work is required to understand the molecular underpinning of this mechanism.   Such understanding of shear-induced disentanglement may come from refinements of slip-link models or tube models or may instead require ``bottom-up'' coarse-graining of atomistic models\cite{baglrv-03,jp-21} that maintain chemical specificity.  Any refinement of the disentanglement mechanism obtained from a greater understanding of this phenomena can readily be expressed in a thermodynamically consistent method through careful specification of the mobility tensors.  The non-equilibrium thermodynamics framework provides a powerful template for modeling the flow of entangled polymer melts in a thermodynamically consistent manner.  

\section*{Acknowledgments}

The authors sincerely thank Marco A. G. Cunha, Mark O. Robbins, Jon E. Seppala, Gretar Tryggvason, and Thao (Vicky) Nguyen, and Daniel Read for useful discussions during the development of the model.  Particular thanks are due to Marco A. G. Cunha and Mark O. Robbins for providing the data from the molecular dynamics simulations.  This work was funded by NSF DMREF 1628974. PDO thanks Georgetown University and the Ives Foundation for support.

\appendix

\section{Reversible dynamics of an entangled polymer melt}

\subsection{Conformation tensor}\label{app:conformation-tensor}
The conformation tensor should encode the stress of the melt, and so is defined to be directly proportional to the elastic stress due to chain stretch and orientation
\begin{equation}
\tens{A} \propto \ens{\sum_{i=1}^{Z_e+1}\vec{Q}_i\vec{T}_i}
\end{equation} 
where $\vec{Q}_i$ is the end-to-end vector of tube segment $i$ as illustrated in figure~\ref{fig:tube}, and $\vec{T}_i$ is the tension on that segment.  We assume that the tube segments are Gaussian subchains, yielding a tension
\begin{equation}
\vec{T}_i = \frac{3k_\textrm{B}T\vec{Q}_i}{N_{e,s}b_K^2},
\end{equation}
where $N_{e,i}$ is the number of Kuhn steps in tube segment $i$, $b_K$ is the length of a Kuhn segment, $T$ is the temperature, and $k_\textrm{B}$ is Boltzmann's constant. The Gaussian assumption is equivalent to assuming an infinitely extensible tube; finite extensibility is treated in Sec.~\ref{subsec:free-energy}.    These assumptions yield a conformation tensor in terms the discrete tube segments,  For analytical calculations, it is more convenient to employ a continuous distribution of tube segments, yielding \eqref{eq:conformation-tensor-definition} in the main text.

\subsection{Kinematics of the conformation tensor and entanglement}\label{app:conformation-entanglements-kinematics}
\new{It is generally assumed that the conformation tensor of an entangled melt exhibits upper-convected behavior under reversible deformations.\cite{oldroyd-50,BerisEdwards,LarsonBook} However, this assumption neglects changes in the conformation tensor due to flow-induced disentanglement.  A reversible deformation does not produce entropy, and hence will not incur dissipation.}  Under such deformation, the bulk material is deformed and the melt may also disentangle or re-entangle at the microscale. \new{As a consequence, the conformation tensor and entanglements are transformed from some reference state to the current state, i.e. $(\tilde{\tens{A}},\tilde\nu)\to(\tens{A},\nu)$}
We thus need two quantities to characterize the deformation: a material deformation tensor encoding the bulk deformation, and a prescription for relabeling the tube segments as entanglements are removed.  We use the deformation tensor
\begin{equation}
	F_{iJ} = \frac{\p x_j}{\p X_J},
\end{equation}
which relates the reference ($\tens{X}$) and current ($\vec{x}$) positions of a material point. Transforming from the reference number of entanglements $(\tilde Z_e)$ to the current $(Z_e)$ relabels the tube segments as
\begin{equation}
	s = \frac{Z_e + 1}{\tilde{Z}_e + 1}\tilde s,
\end{equation}
where $s$ and $\tilde s$ are the current and reference labels of a tube segment.  This expression for re-labeling assume that entanglements are randomly spaced along the chain.

The tube segment vectors $\vec{Q}(s)$ are influenced by both the bulk deformation of the material and the change in the number of entanglements.  A bulk deformation stretches and orientates the tube segment vectors, while adding or removing entanglements splits or joins tube segments.  If we assume \new{that} the tube segments in the reference configuration follow Gaussian statistics, then the newly split and joined segments will also obey Gaussian statistics.  Recalling that $\ens{\vec{Q}(s)\vec{Q}(s)}\sim N_{e}(s)$ for fixed $s$, we find that the tube-segment vectors transform as
\begin{equation}\label{eq:transformation-tube-vector}
	\ens{\frac{\vec{Q}(s)\vec{Q}(s)}{N(s)}} = \ens{\frac{\tens{F}\cdot\tilde{\vec{Q}}\tilde{\vec{Q}}\cdot\tens{F}^\intercal}{\tilde N_e(\tilde s)}},
\end{equation}
under an \new{reversible (and affine)} deformation, where the deformation tensor $\tens{F}$ accounts for bulk material deformation and the transformation $N_e(s)\to\tilde N_e(\tilde s)$ accounts for the change in the average length of a tube vector upon adding or removing entanglements.

The relationship between the reference and current conformation tensor is found by substituting \eqref{eq:transformation-tube-vector} into \eqref{eq:conformation-tensor-definition}, yielding the transformation law
\begin{equation}\label{eq:conformation-tensor-transformation}
	\tens{A} = \frac{Z_{e,eq}\nu + 1}{Z_{e,eq}\tilde{\nu} + 1}\tens{F}\cdot\tilde{\tens{A}}\cdot\tens{F}^\intercal,
\end{equation}
where
\begin{equation}\label{eq:conformation-tensor-reference}
\tilde{\tens{A}} = \frac{3}{b_K^2(Z_{e,eq} + 1)}\ens{\int_0^{\tilde{Z}_e+1}
	\frac{\tilde{\vec{Q}}\tilde{\vec{Q}}}{\tilde{N}_e(\tilde s)} \diff \tilde s}.
\end{equation}
and  we have assume that the ensemble averages of entanglements and conformation are independent when the deformation is affine. 
When the reference state is taken as equilibrium ($\tilde{\tens{A}}=\tens{I},\tilde{\nu}=1$) and the number of entanglements are fixed, the transformation law becomes the Finger deformation tensor
\begin{equation}\label{eq:Finger-tensor}
\tens{A} = \tens{F}\cdot\tens{F}^\intercal
\end{equation}
This implies that any constitutive equation constructed from the conformation tensor \eqref{eq:conformation-tensor-definition} will approach that of an elastic solid in the limit of infinite relaxation time.\cite{sphe-20}

The transformation law \eqref{eq:conformation-tensor-transformation} can be used to find the time derivative of the conformation tensor under reversible flows.  The time derivative of the deformation gradient is
\begin{equation}\label{eq:deformation-reversible}
\frac{\D \tens{F}}{\D t} = \left(\bnabla\vec{v}\right)^\intercal\cdot\tens{F}.
\end{equation}
\new{Differentiating \eqref{eq:conformation-tensor-transformation} with respect to time yields
\begin{equation}
    \left.\mathcal{D}_A\tens{A}\right|_\textrm{rev.} = \tens{0},
\end{equation}
where
\begin{equation}\label{eq:dA-reversible-general}
    \mathcal{D}_A\tens{A}\equiv\stackrel{\triangledown}{\tens{A}} - \frac{Z_{e,eq}}{Z_{e,eq}\nu + 1}\tens{A}\left.\frac{\D \nu}{\D t}\right|_\textrm{rev.} 
\end{equation}
is the convective time derivative of the conformation tensor, which vanishes identically under reversible deformations.}

The reversible time derivative in \eqref{eq:dA-reversible-general} encodes the reduction in stretch from removal of entanglements.  We obtain the reversible time derivative of the entanglements from scaling theories suggest that entanglements are binary interactions between chains, and hence the number of entanglements is proportional to the density of the melt.\cite{milner-20}  From this proportionality, it follows that the entanglements will satisfy a continuity equation in reversible flows, and hence
\new{\begin{equation}\label{eq:dnu-reversible}
    \left.\mathcal{D}_\nu\nu\right|_\textrm{rev.} = 0
\end{equation}
where 
\begin{equation}
    \mathcal{D}_\nu\nu = \frac{\D \nu}{\D t} + \nu\bnabla\cdot\vec{v}
\end{equation}
is the convective derivative of the entanglements.}  Physically, dilation increases the average distance between molecules, and hence decreases the chance that any two molecules will interact.  This in turn decreases the number of entanglements.  Substituting \eqref{eq:dnu-reversible} into \eqref{eq:dA-reversible-general} yields the convective time derivative
\begin{equation}\label{eq:dA-reversible}
	\mathcal{D}_A\tens{A} = \stackrel{\triangledown}{\tens{A}} + \frac{Z_{e,eq}\nu}{Z_{e,eq}\nu + 1}\tens{A}\bnabla\cdot\vec{v}
\end{equation}
The first term on the right-hand side represents affine deformation by the flow and the second term represents a shrinkage of the tube conformation tensor due to the destruction of tube segments in a disentangling melt.  
\new{We will use the reversible dynamics encoded in \eqref{eq:dA-reversible} and \eqref{eq:dnu-reversible} to compute the stress tensor. The elastic stress tensor is obtained from the requirement that the entropy production rate $\dot{S}$ is identically zero for reversible deformation.  It follows that\cite{BerisEdwards,ps-04}
\begin{equation}
    \begin{split} T\dot{S}|_\textrm{rev.} & = \left(\vec{\sigma}:\bnabla\vec{v} + \frac{\delta F}{\delta \tens{A}}:\dot{\tens{A}}|_\textrm{rev.} + \frac{\delta F}{\delta \nu}\dot{\nu}|_\textrm{rev.} \right) \\
    & = 0,
    \end{split}
\end{equation}
which is satisfied by \eqref{eq:stress} in the main text. The dilational terms in the convected derivatives give rise to the isotropic pressure in the stress.  The stress tensor can also be immediately obtained by expressing the reversible dynamics in bracket form.\cite{BerisEdwards}

In the main text, we restrict the convective derivatives \eqref{eq:dA-reversible} and \eqref{eq:dnu-reversible} to incompressible flows so that {$\bnabla\cdot\vec{v}=0$}.  We expect the dilational terms to lead to non-trivial couplings in entangled polymer solutions, whose dynamics can be derived using a two-fluid approach \cite{milner-93}.
}

\subsection{Stretch and orientation}\label{app:stretch-orientation-conformation}
Here, we seek a general relationship between the stretch, orientation, and conformation of an entanglement melt.  We begin by using the results of Sec.~\ref{subsec:conformation} to examine how tube stretch \eqref{eq:stretch-tube} transforms between reference frames, yielding
\begin{equation}\label{eq:stretch-transformation}
\lambda_T = \frac{1}{L_{eq}}\sqrt{\frac{Z_{e,eq}\nu + 1}{Z_{e,eq}\tilde{\nu} + 1}}\ens{\int_0^{\tilde{Z}_e+1}|\tilde{\vec{Q}}\cdot\tens{F}|\diff \tilde s},
\end{equation}
where we have used
\begin{equation}
N(s) = \frac{\tilde Z_{e,eq} + 1}{Z_{e,eq} + 1}\tilde{N}(\tilde s),
\end{equation}
which states that increasing the number of entanglements will decrease the number of Kuhn steps per tube segments.  Taking the time derivative using \eqref{eq:deformation-reversible} and \eqref{eq:dnu-reversible} yields
\begin{equation}
\begin{split}
\frac{\D \lambda_T}{\D t}\rev & = \frac{1}{L_{eq}}\ens{\int_0^{Z_e+1}|\vec{Q}|\frac{\vec{Q}\vec{Q}:\bnabla\vec{v}}{|\vec{Q}|^2}\diff s} \\
& \quad - \frac{1}{2}\frac{Z_{e,eq}\nu\lambda_T}{Z_{e,eq}\nu + 1}\bnabla\cdot\vec{v}, 
\end{split}
\end{equation}
Assuming the averages of stretch and orientation decouple, we arrive at 
\begin{equation}\label{eq:stretch-kinematic}
    \frac{\D \lambda_T}{\D t}\rev = \lambda_T\tens{S}_T:\bnabla\vec{v} -
        \frac{1}{2}\frac{Z_{e,eq}\nu\lambda_T}{Z_{e,eq}\nu + 1}\bnabla\cdot\vec{v},
\end{equation}
where
\begin{equation}
\tens{S}_T = \ens{\frac{\vec{QQ}}{|\vec{Q}|^2}}
\end{equation}
is the average tube orientation tensor. By definition, $\tr\tens{S}_T=1$.

From these transformation laws, one can readily demonstrate that $\tens{S}_T$, $\lambda_T$, and $\tens{A}$ are homogeneous functions of degree zero, one, and two, respectively, in $\tens{F}$ when the number of entanglements are fixed.  We take these as general properties that all definitions of the stretch $\lambda$ and orientation tensor $\tens{S}$ should satisfy.  It follows that $\lambda(\tens{A})$ is a homogeneous function of degree one-half in $\tens{A}$, and $\tens{S}(\tens{A})$ is a homogeneous function of degree zero.  Then, from Euler's theorem for homogeneous functions, we obtain
\begin{subequations}
	\begin{align}
	\tens{A}:\frac{\p }{\p \tens{A}} \lambda&= \frac{1}{2}\lambda, \label{eq:Euler-stretch}\\
	\tens{A}:\frac{\p}{\p \tens{A}}  \tens{S} & = 0. \label{eq:Euler-oreintation}
	\end{align}
\end{subequations}

The kinematics in \eqref{eq:stretch-kinematic} represent a general relationship between stretch and orientation subject to the assumption that stretch and orientation decouple.  A closed-form expression for the orientation tensor can be obtained from these kinematics under the assumption that the orientation tensor is a single-valued function of the conformation tensor.  The time evolution of the stretch and conformation tensor can be related via the chain rule
\begin{equation}\label{eq:stretch-chain-rule}
\frac{\D \lambda}{\D t}  = \frac{\p \lambda}{\p \tens{A}}:\frac{\D \tens{A}}{\D t}.
\end{equation}
We solve \eqref{eq:dA-reversible} for the material derivative of the conformation tensor and substitute into \eqref{eq:stretch-kinematic} using \eqref{eq:stretch-chain-rule} to obtain
\begin{equation}
\begin{split}
\left.\frac{\D \lambda}{\D t} \right|_\textrm{rev.} & = \lambda \left(2\tens{A}\cdot\frac{\p \ln\lambda}{\p \tens{A}}\right):\bnabla\vec{v} \\
&\quad - \frac{1}{2}\frac{Z_{e,eq}\nu\lambda}{Z_{e,eq}\nu + 1}\bnabla\cdot\vec{v},
\end{split}
\end{equation}
where we have used \eqref{eq:Euler-stretch} to simplify the second term.  Comparing with \eqref{eq:stretch-kinematic} yields \eqref{eq:stretch-orientation-equation} in the main text.

\section{Tube stretch}\label{app:tube-stretch}
In general flows, the tube stretch $\lambda_T$ defined in \eqref{eq:stretch-tube} will differ from the stretch $\lambda=\sqrt{\tr\tens{A}/3}.$  Here we \new{identify} approximate bounds on this difference. If the number of entanglements is sufficiently large we may approximate the tube stretch by the Doi-Edwards stretch\cite{DoiEdwards_Book}
\begin{equation}
	\lambda_T\simeq \lambda_{DE} =     \frac{1}{4\pi}\int|\vec{u}\cdot\tens{A}^{1/2}|\diff\vec{u},
\end{equation}
where the integral is taken over the unit sphere.  The Doi-Edwards stretch is bounded from below and above by the stretch $\lambda=\tr{\tens{A}/3}$ as 
\begin{equation}
	\frac{\sqrt{3}}{2}\lambda \leq \lambda_{DE} \leq \lambda.
\end{equation}
The lower bound is obtained in the fully aligned state, where all but one eigenvalues of the conformation tensor vanish, and the upper-bound corresponds to an isotropic melt.  That is, the stretch \new{measure used in our model (and the Rolie-Poly model)} is identical to the Doi-Edwards stretch in the isotropic case, and over predicts the Doi-Edwards stretch in the fully aligned case.

\section{Ianniruberto-Marrucci stretch and orientation}\label{app:im}
The orientation tensor $\tens{S}$ in the I-M model is computed via a history integral
\begin{equation}\label{eq:orientation-CCR}
\tens{S}(t) = \int_{-\infty}^t\frac{1}{\tau_d(t^\prime)}
\exp\left(\int_{t^\prime}^t \frac{\diff t^{\prime\prime}}{\tau_d(t^{\prime\prime})}\right)\tens{S}_\textrm{rev.}(t,t^\prime) \diff t^\prime,
\end{equation}
where the relaxation rate
\begin{equation}\label{eq:im-reptation-time}
\frac{1}{\tau_d(t)} = \frac{1}{\tau_{d,eq}} + \beta\left(\tens{S}:\bnabla\vec{v} - \frac{1}{\lambda}\frac{\D \lambda}{\D t}\right)
\end{equation}
depends on the current tube orientation.  The reversible orientation tensor $\tens{S}_\textrm{rev.}(t,t^\prime)$ is the orientation tensor in the absence of relaxation $(\tau_d(t)\to\infty)$, and depends on $\tens{F}(t,t^\prime)$, the relative deformation tensor between $t^\prime$ and $t$.  Originally, \citet{im-14} computed $\tens{S}_\textrm{rev.}$ from the Doi-Edwards alignment tensor.  However, \citet{ianniruberto-15} demonstrated that the Seth tensor
\begin{equation}\label{eq:Q}
\tens{S}_\textrm{rev.}(t,t^\prime) = \frac{(\tens{F}(t,t^\prime)\cdot\tens{F}^\intercal(t,t^\prime))^{1/3}}{\tr (\tens{F}(t,t^\prime)\cdot\tens{F}^\intercal(t,t^\prime))^{1/3}}
\end{equation}
is better able to describe transient relaxation of the shear stress following a step strain.  I-M assumed that the stretch relaxes on the Rouse time according to
\begin{equation}\label{eq:im-stretch-relaxation}
\frac{\D \lambda}{\D t}  = \lambda\tens{S}:\bnabla\vec{v} - \frac{\lambda - \nu^{1/2}}{\tau_R},
\end{equation}
where the reference stretch $\nu^{1/2}$ arises from the assumption that the force on a chain decreases with increasing tube diameter. As a consequence, the stretch in the I-M model relaxes towards a non-equilibrium tube of length  \new{$L\sim \nu^{1/2}$} on the Rouse time.  

\section{Stretch and orientation dynamics}\label{app:stretch-orientation}
\begin{widetext}
The governing equations for $\tens{A}$ can be re-expressed in terms of the orientation tensor $\tens{S}$ and stretch $\lambda$ using the chain rule:
\begin{subequations}\label{eq:governing-equations-decouple}
		\begin{align}
		\begin{split}
		\stackrel{\triangledown}{\tens{S}}  + 2\tens{SS}:\bnabla\vec{v} & = 
		- \frac{1}{\lambda^2}\left(\frac{\tens{I}}{\tau_d(\lambda)} + \alpha\frac{3\lambda^2\tens{S}-\tens{I}}{\tau_d(\lambda)}\right)\cdot\left(\tens{S} - \frac{1}{3}\tens{I}\right) 
		+ \frac{3\alpha}{\tau_d(\lambda)}\tens{S}\tens{S}:\left(\tens{S} - \frac{1}{3}\tens{I}\right) \\
		& \quad - \frac{3\alpha}{\tau_d(\lambda)}\left(\tens{S}^2 - \tens{S}\tr\tens{S}^2\right)\left(\f(\lambda)\lambda^2 - 1\right) - \frac{\alpha\zeta_Z\beta\nu}{\tau_d(\lambda)}\left(\tens{S}^2 - \tens{S}\tr\tens{S}^2\right)\ln\nu, 
		\end{split}\label{eq:governing-equation-S}\\
		\begin{split}
		\frac{\D \lambda}{\D t}  - \lambda\tens{S}:\bnabla\vec{v} & = 
		- \frac{3\alpha\lambda}{2\tau_d(\lambda)}\tens{S}:(\tens{S} - \frac{1}{3}\tens{I}) 
		- \frac{1 + \alpha(3\lambda^2\tr\tens{S}^2 - 1)}{2\tau_d(\lambda)\lambda} (\lambda^2\f(\lambda) -1) 
		- \frac{1}{\tau_R}\frac{\lambda^2\f(\lambda) -1}{\lambda+1}\\
		& \quad - \frac{\zeta_Z\beta\nu}{6\lambda}
		\left(\frac{\tens{I}}{\tau_d(\lambda)} + \alpha\frac{3\lambda^2\tens{S} - \tens{I}}{\tau_d(\lambda)} 
		+ \frac{2\lambda \tens{I}}{\tau_R(\lambda + 1)}\right):\tens{S}\ln\nu. 
		\end{split}\label{eq:governing-equation-lambda}
		\end{align}
	\end{subequations}
\end{widetext}

\section{Green-Kubo relations for the mobility tensor}\label{app:green-kubo}
The conformation tensor fluctuates rapidly on the Rouse timescale $\tau_e$ of an entanglement segment.  On longer times, motion of the conformation tensor is deterministic (for our description), with fluctuations leading to dissipation through the Green-Kubo relation\cite{Ottinger_Book} 
\begin{equation}\label{eq:green-kubo}
\mathbb{M}^{AA} = \frac{1}{nk_\textrm{B}T}\ens{\int_0^{\infty}\dot{\tens{A}}^\textrm{dis.}(t)\dot{\tens{A}}^\textrm{dis.}(t+\bar{t}) \diff \bar{t}}.
\end{equation}
The over dot indicates time differentiation and the superscript ``$\textrm{dis.}$'' indicates that only dissipative (irreversible) changes are included.  The noise occurs on timescales faster than $\tau_e$.  A segmental description of the melt is required for shorter timescales.  

The dissipative time derivative of the tube-segment vectors is defined as the total rate minus the affine rate:
\begin{equation}
	\dot{\vec{Q}}^\textrm{dis}(s) \equiv  \dot{\vec{Q}}(s) - \vec{Q}(s)\cdot\bnabla\vec{v}.
\end{equation}
We assume that fluctuations in the tube-segment length are additive, such that the dissipative time derivative can be expressed as
\begin{equation}\label{eq:bond-vector-equation}
\dot{\vec{Q}}^\textrm{dis}(s)  = \vec{W}^\textrm{rep} + \vec{W}^\textrm{ret},
\end{equation}
where the dissipative random vectors $\vec{W}^\textrm{rep}$ and $\vec{W}^\textrm{ret}$ encode reptation and retraction, respectively. The separation of degrees of freedom faster than $\tau_e$ into processes representing what are eventually slower degrees of freedom (reptation and retraction) is a non-trivial and strictly non-justified assumption. We make this choice here in order to obtain a correspondence with the \new{I-M and Rolie-Poly models}. It is possible that these two forms of noise are loosely identified with intrachain fluctuations (retraction) and inter-chain fluctuations (reptation), but this is surely incorrect in detail. We regard this as an open challenge.

\begin{widetext}
These vectors have zero mean and units of velocity.  Differentiating \eqref{eq:conformation-tensor-definition} and substituting \eqref{eq:bond-vector-equation} yields
\begin{equation}\label{eq:conformation-tensor-fluctutations}
	\dot{A}_{ij}^\textrm{dis.} = \frac{3}{b_K^2( Z_{e,eq} + 1)}\int_0^{Z_e + 1}\frac{1}{N_e(s)}\left[Q_i(W_j^\textrm{rep}(s,t)  + W_j^\textrm{ret}(s,t))  + (W_i^\textrm{rep}(s,t)  + W_i^\textrm{ret}(s,t))Q_j \right] \diff s.
\end{equation}
We do not explicitly include changes in the conformation tensor arising from fluctuations in entanglements; we follow the GLaMM model in including these fluctuations in the dynamics of the tube-segment vectors.\cite{glmm-03}  We assume that there are no cross-correlations between reptation and retraction noises, \textit{i.e.}
\begin{equation}
	\ens{\vec{W}^\textrm{rep}(s,t)\vec{W}^\textrm{ret}(\bar{t},\bar s) } = \tens{0},
\end{equation}
This separation of the noise according to \eqref{eq:bond-vector-equation} allows us to express the conformation tensor mobility as the sum of the effects of reptation and retraction:
\begin{equation}
	\mathbb{M}^{AA} = \mathbb{M}^{AA,\textrm{rep}} + \mathbb{M}^{AA,\textrm{ret}}.
\end{equation}

The reptation noise variance can be written as\cite{glmm-03}
\begin{equation}
	\vec{W}^\textrm{rep}(s,t) = \frac{\p }{\p s}\vec{R}(s+\Delta\xi(t),t),
\end{equation}
where $\Delta\xi(t)$ represents the stochastic re-labeling of tube segments due to reptation.  The reptation mobility can be determined by Fourier transforming the tube coordinates $s$ and $\bar s$ and taking a single-mode approximation.  We take an equivalent approach, and assume the variance of the reptation noise has the form
\begin{equation}\label{eq:reptation-auto-correlation}
\ens{\vec{W}^\textrm{rep}(s,t)\vec{W}^\textrm{rep}(\bar{t},\bar s) }  = \frac{b_K^2(Z_{e,eq} + 1)}{3}
\times\frac{nk_\textrm{B}TN_e(s)}{2}\delta(t-\bar{t})\delta(s-\bar s)\tens{M}^\textrm{rep}(\tens{A}),
\end{equation}
where $\tens{M}^\textrm{rep}$, the mobility tensor for the tube with the equilibrium number of entanglements, is assumed to depend on the average conformation tensor. The prefactor $b_K^2(Z_{e,eq} + 1)/3$ is required for dimensional consistency and the factor of $N_e$ accounts for the increase in chain mobility as the melt disentangles.  Substituting \eqref{eq:conformation-tensor-fluctutations} into \eqref{eq:green-kubo} and averaging using \eqref{eq:reptation-auto-correlation} and \eqref{eq:conformation-tensor-definition}, the definition of the conformation tensor, yields the contribution of reptation to the mobility tensor:
\begin{equation}\label{eq:mobility-reptation}
		M^{AA,\textrm{rep}}_{ijk\ell} = \frac{1}{2}\left(A_{ik}M_{j\ell}^\textrm{rep} + A_{jk}M_{i\ell}^\textrm{rep}
			+ A_{i\ell}M_{jk}^\textrm{rep} + A_{j\ell}M_{ik}^\textrm{rep}\right).
\end{equation}
We complete the reptation mobility by assuming the Giesekus form,  \eqref{eq:Giesekus} for $\tens{M}^{\textrm{rep}}$.

We obtain the retraction mobility by assuming that retraction only acts on the stretch, leaving the orientation tensor $\tens{S}$ unchanged.  A straight-forward application of the chain rule shows that these assumptions yields a retraction mobility of the form
\begin{equation}\label{eq:mobility-Rouse}
    \begin{aligned}
        M_{ijk\ell}^{AA,\textrm{ret}} & = \frac{\p A_{ij}}{\p \lambda}\frac{\p A_{k\ell}}{\p \lambda}M^\textrm{ret},\\
        & = \frac{4}{\lambda^2}A_{ij}A_{k\ell}M^\textrm{ret},
    \end{aligned}
\end{equation}
where the scalar retraction mobility is
	\begin{equation}
	M^\textrm{ret} = \left(\frac{3}{b_K^2( Z_{e,eq} + 1)}\right)^2\frac{1}{9nk_\textrm{B}T\lambda^2}\int_0^\infty\ens{\int_0^{1+Z_e}\int_0^{1+Z_e}\frac{(\vec{Q}(s)\cdot\vec{W}^\textrm{ret}(s,t))  (\vec{Q}(\bar s)\cdot\vec{W}^\textrm{ret}(\bar s,t))}
		{N_e(s)N_e(\bar s)}\diff s\diff \bar s}\diff t.
	\end{equation}
By making a suitable choice for the retraction noise we can recover the form of the mobility implicitly used in the Rolie-Poly model, \eqref{eq:rolie-poly}.

Tube theories model contour length fluctuations with a stochastic force that is localized to individual tube segments.\cite{lm-02} Here, we take a similar approach and assume a retraction noise of the form 
\begin{equation}\label{eq:retraction-auto-correlation}
\ens{\vec{W}^\textrm{ret}(s,t)\vec{W}^\textrm{ret}(\bar s,\bar{t}) }     = nk_\textrm{B}TN_e(s)b_K^2(Z_{e,eq} + 1)
  \delta(t-\bar{t})\delta(s-\bar s)M^\textrm{ret}(\tens{A})\frac{\vec{Q}(s)\vec{Q}(\bar s)}{|\vec{Q}(s)||\vec{Q}(\bar s)|},
\end{equation}
where the factors of $\vec{Q}$ project the retraction noise in the direction along the chain.
This (non-unique) choice of noise leads to \eqref{eq:rolie-poly}. Combining \eqref{eq:mobility-Rouse} with \eqref{eq:mobility-reptation} yields \eqref{eq:MAA-general} in the main text.

\end{widetext}

\bibliography{}{}

\end{document}